%% file: Audio_Inpainting4arxiv.tex
\documentclass[preprint,12pt]{elsarticle}

\usepackage{amssymb}
\usepackage{amsfonts}
\usepackage{amsmath}
\usepackage{dsfont}

\usepackage{url}
\usepackage{comment}
\usepackage{algorithm}
\usepackage[noend]{algpseudocode}
\usepackage{graphicx}
\usepackage{comment}
\usepackage{balance}
\usepackage{acronym}
\usepackage{amsmath,amsfonts}
\usepackage{amssymb}
\usepackage{tabularx}
\usepackage[dvipsnames]{xcolor}
\usepackage{subcaption}
\usepackage{booktabs}
\usepackage{multirow}
\usepackage{bbm}
\usepackage{mathtools}
\usepackage{hyperref}

\input{definitions}
\input{acronyms}

\journal{Speech Communication}

\begin{document}

\begin{frontmatter}



\title{Transient Noise Removal via Diffusion-based Speech Inpainting}


\author{Mordehay Moradi and Sharon Gannot} 

\affiliation{organization={Faculty of Engineering, Bar-Ilan University},
            city={Ramat-Gan},
            postcode={5290002}, 
            country={Israel}
            }

\newpage
\begin{abstract}
\label{abstract}
In this paper, we present \ac{PGDI}, a diffusion-based speech inpainting framework for restoring missing or severely corrupted speech segments. Unlike previous methods that struggle with speaker variability or long gap lengths, \ac{PGDI} can accurately reconstruct gaps of up to one second in length while preserving speaker identity, prosody, and environmental factors such as reverberation. Central to this approach is classifier guidance, specifically phoneme-level guidance, which substantially improves reconstruction fidelity. \ac{PGDI} operates in a speaker-independent manner and maintains robustness even when long segments are completely masked by strong transient noise, making it well-suited for real-world applications, such as fireworks, door slams, hammer strikes, and construction noise. Through extensive experiments across diverse speakers and gap lengths, we demonstrate \ac{PGDI}’s superior inpainting performance and its ability to handle challenging acoustic conditions. We consider both scenarios, with and without access to the transcript during inference, showing that while the availability of text further enhances performance, the model remains effective even in its absence.
For audio samples, visit: \url{https://mordehaym.github.io/PGDI/}.
\end{abstract}



\end{frontmatter}



\section{Introduction}
Speech inpainting aims to restore missing or severely corrupted speech segments, which are often obscured by severe noise, such as fireworks, door slams, hammer strikes, and construction noise. This task is analogous to image inpainting \cite{image_inpaaingintg}, where missing pixel regions are inferred from the surrounding image content. In speech inpainting, generating natural-sounding reconstructions requires preserving environmental conditions---such as reverberation and background noise---as well as maintaining speaker identity and speech prosody.

Unlike prior works on packet loss concealment \cite{bs_plc, complex_bin2bin}, which typically address missing regions at the sample level or short gaps of up to 250 ms, speech inpainting operates at the frame level and can involve gaps spanning multiple seconds. Several studies have explored this problem. For instance, \cite{fill_in_gap} examines the application of \ac{SSL} models to reconstruct missing speech segments based on their surrounding context. Similarly, \cite{lstm_si} employs multi-layer \ac{LSTM} networks for speech inpainting, demonstrating high perceptual quality for gaps up to one second in single-speaker scenarios.  

However, these approaches exhibit limitations when dealing with multi-speaker scenarios or gaps exceeding 400 ms, as they rely solely on audio-based information. For gaps longer than 1000 ms, the missing regions may encompass multiple words, making it difficult to reconstruct the intended speech content without additional contextual cues. To address this challenge, some works integrate textual information into the inpainting process \cite{voice_conversion_2016, borsos2022speechpainter}.  

The approach in \cite{voice_conversion_2016} performs text-informed speech inpainting through voice conversion, considering gaps of up to 750 ms. It relies on a \ac{GMM} trained on a parallel dataset for each speaker, where both source and target speakers utter identical linguistic content, enabling precise alignment of their speech features. This requirement poses challenges in scenarios where collecting large-scale speaker-specific data is impractical. 
Another text-based method is proposed in \cite{borsos2022speechpainter}, where textual information is integrated into the Perceiver IO model \cite{precevierio} for speech inpainting. However, this approach is limited to 3-second input segments because it relies on learned queries tied to a fixed duration. Extending it to longer signals requires either a chunking strategy or retraining on longer-duration data. Moreover, the method is deterministic, which restricts the diversity of the generated outputs. A common limitation of such text-based methods \cite{borsos2022speechpainter, voice_conversion_2016} is the need to provide the corresponding text at inference time.

These limitations highlight the need for more flexible and robust speech inpainting methods that can handle longer gaps and multi-speaker scenarios, generate diverse outputs, and produce high-quality reconstructions, while remaining effective even without access to the transcript at inference time, and further benefiting from it when available.

Diffusion models have recently emerged as a powerful generative approach, demonstrating state-of-the-art performance in tasks such as image synthesis \cite{ho2020denoising}, speech generation \cite{grad-tts}, and audio restoration \cite{audio_restoration}. These models learn to iteratively refine noisy signals toward realistic outputs by reversing a predefined noise-corruption process. For speech inpainting, diffusion-based models offer several advantages: they can generate high-quality, coherent speech reconstructions, leverage strong probabilistic modeling, and flexibly handle different durations of missing speech.  

In diffusion models, two main strategies are used to direct the generation process: 1) classifier guidance \cite{general_classifier_guidance} and 2) classifier-free guidance \cite{classifier_free}. Classifier guidance relies on a separately trained classifier to steer the generation toward more realistic speech samples, offering greater flexibility in shaping the reconstructed output. However, this approach requires an additional classification model, increasing computational complexity. In contrast, classifier-free guidance conditions the generation process directly on relevant information, such as speaker identity and phonetic context, without needing an external classifier. This method simplifies the pipeline while allowing control over the inpainted speech.  
Both approaches have been successfully applied in generative modeling, and their effectiveness in speech inpainting depends on the trade-off between computational efficiency and reconstruction quality.  

In this work, we introduce \ac{PGDI}, a novel approach that leverages diffusion models for speech inpainting by exploiting their ability to generate realistic speech conditioned on the surrounding context. \ac{PGDI} reconstructs missing speech segments through a progressive denoising process, enabling smooth transitions and high-fidelity synthesis. We further guide the process with text-based information extracted using a language model, thereby improving the quality and accuracy of the reconstruction. 
Our findings demonstrate the effectiveness of \ac{PGDI} across a wide range of scenarios, from short gaps to long missing segments, making it particularly suitable for reconstructing speech corrupted by severe distortions, such as strong and abrupt noise. We evaluate the method in both settings, with and without access to the ground-truth transcript during inference, and observe that while the availability of text enables strong performance even for long gaps, \ac{PGDI} without text remains effective for relatively shorter gaps. 

The method offers several key advantages: 1) it is speaker-independent, requiring no prior knowledge of the speaker’s identity, 2) it preserves the speaker’s natural voice, style, and prosody, and 3) it maintains environmental consistency, including reverberation characteristics. These advantages are valid regardless of the availability of the transcript during inference.


The paper is organized as follows. Section~\ref{sec:problem} defines the core problem and outlines the task setting. Section~\ref{sec:method} presents the proposed method, including the inference-time data pipeline and the design of the model architectures. Section~\ref{sec:experiments} details the experimental setup, evaluation metrics, baseline comparisons, and results. Finally, Section~\ref{sec:conclusion} summarizes the key findings and discusses potential directions for future work.

\section{problem formulation}
\label{sec:problem}

Let $\mathbf{x} \in \mathbb{R}^d$ denote the speech signal represented as a vectorized mel-spectrogram, where $d = F \times L$ is the total number of frequency-time elements, with $F$ representing the number of mel-frequency bins and $L$ the number of time-frames.

We define a binary mask $\mathbf{m} \in \{0,1\}^d$, where $m_i = 1$ indicates that the $i$-th element of the spectrogram is observed, and $m_i = 0$ indicates that it is masked or missing. The mask $\mathbf{m}$ can be either provided by the user or generated by an external module that detects frames affected by high levels of noise. In this study, the mask $\mathbf{m}$ is structured such that the same binary value is applied uniformly across all frequency bins within each time frame; hence, each frame is either fully observed or fully masked.

The masked spectrogram is obtained by the Hadamard (element-wise) product of the mask and the original signal:
\begin{equation}
    \mathbf{x}_\mathrm{m} = \mathbf{m} \odot \mathbf{x}.
\end{equation}
Given the masked mel-spectrogram and the mask, the goal is to reconstruct the missing frames such that the observed frames remain unaltered, while the masked frames are plausibly inpainted to match the unmasked frames in terms of perceptual consistency, capturing the correct prosody, speaker identity, and acoustic characteristics.

\section{Proposed Method}
\label{sec:method}
Our method employs a dual-guidance strategy, combining text-based classifier guidance (inferred from the available masked signal) and classifier-free guidance to strike a balance between semantic fidelity and variability of the generated speech. Text-based guidance aligns the generated speech with linguistic content, while classifier-free guidance enhances variability in acoustic characteristics.

\subsection{Data Pipeline at the Inference Stage}
\label{method:inference}
In this section, we describe the inference pipeline of the proposed \ac{PGDI} model. The proposed \ac{PGDI} model operates during the inference phase by incorporating a text-based module, specifically a phoneme classifier, into a diffusion-based generative process. The goal is to generate high-quality mel-spectrograms corresponding to intelligible speech that is aligned with a target sentence.

\begin{figure*}[htbp]
    \centering
      \includegraphics[width=0.95\textwidth]{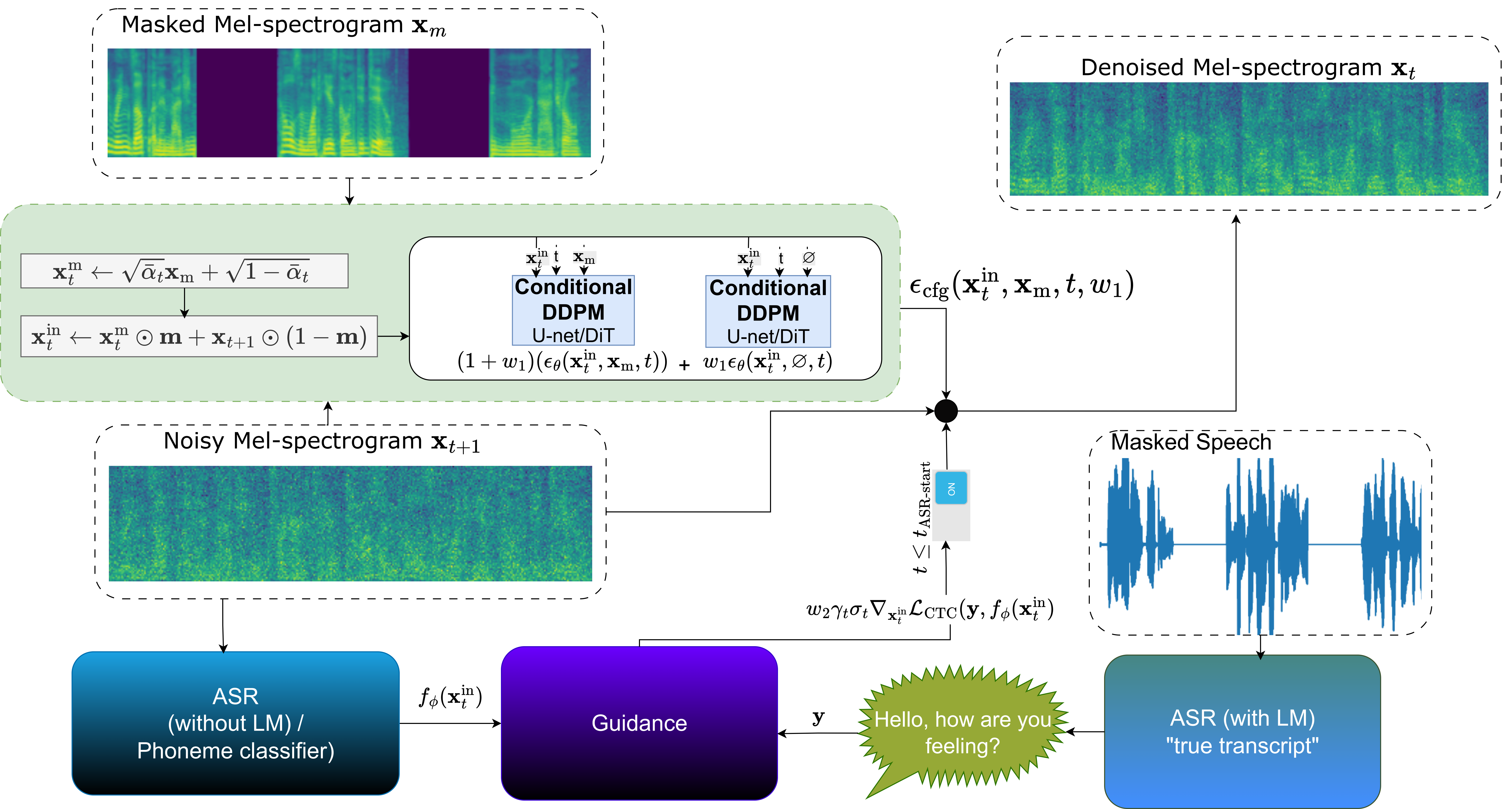}
      \setlength{\belowcaptionskip}{0pt}
      \caption{High-level block diagram of the proposed model at the inference phase. A time-domain reconstruction module is used but not shown.}
\label{fig:model_arch_high_level_inference}
\end{figure*}

As shown in Fig.~\ref{fig:model_arch_high_level_inference}, the overall process includes several key modules for inpainting the signal. The masked speech signal is initially processed by an \ac{ASR} system, with incorporated \ac{LM}. This ASR+LM module is used once per utterance to predict the desired text. We employ the system from \cite{inference_asr_lm}, which is chosen for its low \ac{WER}. It achieves a \ac{WER} of 1\% and 1.5\% on the LRS3 \cite{LRS3} and LRS2 \cite{LRS2} datasets, respectively. As discussed in the sequel, the incorporated \ac{LM} helps recover missing or obscured words in the input, enabling the generated transcription to serve as the desired target sentence to guide the audio generation process.
To synthesize the denoised mel-spectrogram, we use our trained conditional DDPM, which operates iteratively to progressively refine the mel-spectrogram at each step, with model weights $\boldsymbol{\theta}$.

At each reverse step $ t $, where $ t $ decreases from $ T-1 $ to 0 (with $ T $ the total number of timesteps in the reverse process), a noisy version of the masked mel-spectrogram is generated. The noisy signal at step $ t $, denoted as $ \mathbf{x}^\mathrm{m}_t $, is obtained by combining the clean masked mel-spectrogram $ \mathbf{x}_\mathrm{m} $ with Gaussian noise $ \boldsymbol{\epsilon} \sim \mathcal{N}(0, \mathbf{I}) $, where $ \mathbf{I} \in \mathbb{R}^{d \times d} $ is the identity matrix, according to the diffusion equation:
\begin{equation}
\label{eq:noisnig_inference}
\mathbf{x}^\mathrm{m}_t \leftarrow \sqrt{\bar{\alpha}_t} \mathbf{x}_\mathrm{m} + \sqrt{1-\bar{\alpha}_t} \boldsymbol{\epsilon}.
\end{equation}
Inspired by~\cite{repaint2022},  we define the input to the DDPM at step $ t $ as $ \mathbf{x}_t^{\textrm{in}} $.  It is computed by combining the noisy masked mel-spectrogram $ \mathbf{x}^\mathrm{m}_t $ and the current estimate $ \mathbf{x}_{t+1} $, according to the mask structure:
\begin{equation}
\label{eq:input2inference}
\mathbf{x}_t^{\textrm{in}} \leftarrow \mathbf{x}^\mathrm{m}_t \odot \mathbf{m} + \mathbf{x}_{t+1} \odot (1-\mathbf{m}).
\end{equation}
This step ensures that noise is appropriately added to $ \mathbf{x}^\mathrm{m}_t $ according to the diffusion schedule, preserving information from the known regions of $ \mathbf{x}_\mathrm{m} $ while introducing stochasticity into the masked regions.

The initial signal $ \mathbf{x}_T $ is sampled from a standard Gaussian distribution, i.e., $ \mathbf{x}_T \sim \mathcal{N}(0, \mathbf{I}) $, hence, the denoising process starts from pure noise. 

After preparing the input $ \mathbf{x}_t^{\textrm{in}}  $ of the DDPM, the classifier-free guidance mechanism is applied to produce the noise estimate, combining conditional and unconditional predictions:
\begin{equation}
    \boldsymbol{\epsilon}_{\text{cfg}}(\mathbf{x}_t^{\textrm{in}} , \mathbf{x}_\mathrm{m}, t, w_1)  = (1 + w_1) \boldsymbol{\epsilon}_{\boldsymbol{\theta}}(\mathbf{x}_t^{\textrm{in}} , \mathbf{x}_\mathrm{m}, t) - w_1 \boldsymbol{\epsilon}_{\boldsymbol{\theta}}(\mathbf{x}_t^{\textrm{in}} , \varnothing, t).
    \label{eq:classifier_free}
\end{equation} 
Here, the hyperparameter $ w_1 $ controls the strength of the guidance, and $\mathbf{x}_\mathrm{m} $ serves as the conditioning used in the classifier-free guidance method. In this setting, when conditioning is omitted, it is replaced with a null input denoted by $ \varnothing $.

Subsequently, classifier-based guidance is employed. In this stage, the current noisy mel-spectrogram $ \mathbf{x}_t^{\textrm{in}}  $ is processed by an \ac{ASR} system (without language modeling) at each diffusion step. This \ac{ASR}, trained on noisy mel-spectrograms, generates a transcription (or, alternatively, a phoneme sequence) based solely on the acoustic features of the input.

The alignment between the target sequence and the current \ac{ASR} output is quantified using the \ac{CTC} loss \cite{ctcloss}:
\begin{multline}
\mathcal{L}_{\text{CTC}} = \text{CTC}(
\underbrace{\text{Target Transcription (ASR w.~LM)}}_{\mathbf{y}}, \\
\underbrace{\text{Frame-wise Probabilities (ASR wo.~LM)}}_{\mathclap{f_{\boldsymbol{\phi}}(\mathbf{x}_t^{\textrm{in}} )}})
\label{eq:ctc}
\end{multline}
which measures the mismatch between the two sequences. The target sequence $\mathbf{y}$ consists of a sequence of discrete token indices corresponding to the desired transcription output. This loss guides the refinement of the mel-spectrogram at each denoising step in the DDPM model.

We denote the \ac{ASR} model as $ f_{\boldsymbol{\phi}} $, where $ \boldsymbol{\phi} $ represents the model parameters. Given the current noisy mel-spectrogram $ \mathbf{x}_t^{\textrm{in}}   $, the \ac{ASR} model predicts frame-wise probabilities $ f_{\boldsymbol{\phi}}(\mathbf{x}_t^{\textrm{in}}  ) $. The classifier log-probability is then defined as:
\begin{equation}
\log p_{\boldsymbol{\phi}}(\mathbf{y} \mid \mathbf{x}_t^{\textrm{in}}  ) = -\mathcal{L}_{\text{CTC}}(\mathbf{y}, f_{\boldsymbol{\phi}}(\mathbf{x}_t^{\textrm{in}}  )).
\end{equation}
The gradient of the \ac{CTC} loss is then used to guide the refinement of the noisy mel-spectrogram towards semantically coherent outputs:
\begin{equation}
\hat{\boldsymbol{\epsilon}}(\mathbf{x}_t^{\textrm{in}}  , \mathbf{y}, \mathbf{x}_\mathrm{m}, t, w_1, w_2) = \boldsymbol{\epsilon}_{\text{cfg}}(\mathbf{x}_t^{\textrm{in}} , \mathbf{x}_\mathrm{m}, t, w_1) - w_2 \gamma_t \sigma_t \nabla_{\mathbf{x}_t^{\textrm{in}}  } \log p_{\boldsymbol{\phi}}(\mathbf{y} \mid \mathbf{x}_t^{\textrm{in}}  ),
\label{eq:mg_p_asr}
\end{equation}
where $ w_2 $ controls the strength of the ASR-based guidance, and $ \sigma_t $ is the noise level at timestep $ t $.
The ASR-based guidance is applied only at the last $ t_{\text{ASR-start}} $ steps of the reverse process, allowing the model to first focus on general denoising before enforcing semantic consistency.
At each timestep, after predicting the noise, the next state $ \mathbf{x}_{t} $ is estimated using the sampling equation:
\begin{equation}
\mathbf{x}_{t} \gets \frac{1}{\sqrt{\alpha_t}} \left( \mathbf{x}_t^{\textrm{in}}   - \frac{1 - \alpha_t}{\sqrt{1 - \bar{\alpha}_t}} \, \boldsymbol{\epsilon}_{\text{p}} \right)
+ \mathds{1}_{\{t > 0\}} \cdot \sigma_t \mathbf{z},
\label{eq:update_step}
\end{equation}
where $ \boldsymbol{\epsilon}_{\text{p}} $ denotes either $ \boldsymbol{\epsilon}_{\text{cfg}} $ or $ \hat{\boldsymbol{\epsilon}} $, depending on whether ASR guidance is applied, and $\mathds{1}_{\{t > 0\}}$ is an indicator function that equals 1 when $t > 0$ and 0 otherwise, ensuring no noise is added at the final step. The variable $ \mathbf{z} \sim \mathcal{N}(0, \mathbf{I}) $ is a standard Gaussian noise vector, introducing stochasticity during the sampling process. At the final step $ t = 0 $, we apply the update 
\begin{equation} 
\mathbf{x}_{t=0} \gets \mathbf{x}_\mathrm{m} + \mathbf{x}_{t=0} \odot (1 - \mathbf{m})
\label{eq:final_Step_inference}
\end{equation}to ensure that the unmasked regions remain intact.

During the generation process, it was observed that $ \boldsymbol{\epsilon}_{\text{cfg}} $ was much larger in magnitude compared to $ \nabla_{\mathbf{x}_t^{\textrm{in}} } \log p_{\boldsymbol{\phi}}(\mathbf{y} \mid \mathbf{x}_t^{\textrm{in}} ) $, making it challenging to set $ w_2 $ correctly and resulting in suboptimal mel-spectrogram estimates. To address this, following~\cite{kim2022guided}, a gradient normalization factor was introduced:
\begin{equation}
\gamma_t = \frac{\sqrt{|\boldsymbol{\epsilon}_{\text{cfg}}|}}{\sigma_t \cdot |\nabla_{\mathbf{x}_t^{\textrm{in}} } \log p_{\boldsymbol{\phi}}(\mathbf{y} \mid \mathbf{x}_t^{\textrm{in}}  )|},
\label{eq:grad_norm}
\end{equation}
where $ \| \cdot \| $ denotes the Frobenius norm.

The complete procedure, including both the conditioning and subsequent guidance steps, is detailed in Algorithm~\ref{algorithm_difusion_guidance}. 
By integrating the text-based classifier into the generation process, the model ensures that the final reconstructed mel-spectrogram is both acoustically faithful and semantically meaningful, enhancing the quality and intelligibility of the generated speech in noisy or challenging conditions.

\begin{algorithm}[htbp]
\caption{Diffusion sampling with ASR guidance at inference time.}
\begin{algorithmic}[1]
\State \textbf{Input:} masked audio, masked mel-spectrogram $\mathbf{x}_\mathrm{m}$, mask $\mathbf{m}$, DDPM's weights $\boldsymbol{\theta}$, ASR/Phoneme model's weights $\boldsymbol{\phi}$, diffusion parameters $\{\alpha_t, \bar{\alpha}_t, \sigma_t\}$, guidance weights $w_1, w_2$
\State $\mathbf{y} \gets \text{ASR+LM}(\text{masked audio})$
\State Initialize 
$ \mathbf{x}_T \sim \mathcal{N}(0, \mathbf{I})$
\For{$t = T-1, T-2, \dotsc, 0$}
    \State Sample $\boldsymbol{\epsilon} \sim \mathcal{N}(\mathbf{0}, \mathbf{I})$
    \State $\mathbf{x}^{\mathrm{m}}_t \gets \sqrt{\bar{\alpha}_t} \mathbf{x}_\mathrm{m} + \sqrt{1-\bar{\alpha}_t} \boldsymbol{\epsilon}$ \Comment{Eq.~\eqref{eq:noisnig_inference}}
    \State $\mathbf{x}_t^{\textrm{in}}  \gets \mathbf{x}^\mathrm{m}_t  \odot \mathbf{m} + \mathbf{x}_{t+1} \odot (1 - \mathbf{m})$ \Comment{Eq.~\eqref{eq:input2inference}}
    \State $     \boldsymbol{\epsilon}_{\text{cfg}} \gets (1 + w_1)  \boldsymbol{\epsilon}_{\boldsymbol{\theta}}(\mathbf{x}_t^{\textrm{in}} , \mathbf{x}_\mathrm{m}, t) - w_1 \boldsymbol{\epsilon}_{\boldsymbol{\theta}}(\mathbf{x}_t^{\textrm{in}} , \varnothing, t).$ \Comment{Eq.~\eqref{eq:classifier_free}}

    \If{$t \leq t_{\text{ASR-start}}$}
        \State Compute $\mathcal{L}_{\text{CTC}}(\mathbf{y}, f_{\boldsymbol{\phi}}(\mathbf{x}_t^{\textrm{in}}  ))$ \Comment{Eq.~\eqref{eq:ctc}}
        \State $\nabla_{\mathbf{x}_t^{\textrm{in}} } \mathcal{L}_{\text{CTC}}(\mathbf{y}, f_{\boldsymbol{\phi}}(\mathbf{x}_t^{\textrm{in}}  )) \gets$ gradient of ASR loss w.r.t.\ $\mathbf{x}_t^{\textrm{in}} $
        \State $\gamma_t \gets \frac{\sqrt{\|\boldsymbol{\epsilon}_{\text{cfg}}\|}}{\sigma_t \cdot \|\nabla_{\mathbf{x}_t^{\textrm{in}}} \mathcal{L}_{\text{CTC}}(\mathbf{y}, f_{\boldsymbol{\phi}}(\mathbf{x}_t^{\textrm{in}}  ))\|}$ \Comment{Eq.~\eqref{eq:grad_norm}}
        \State $\hat{\boldsymbol{\epsilon}} \gets \boldsymbol{\epsilon}_{\text{cfg}} - w_2 \gamma_t \sigma_t \nabla_{\mathbf{x}_t^{\textrm{in}}  } \log p_{\boldsymbol{\phi}}(\mathbf{y} \mid \mathbf{x}_t^{\textrm{in}}  ),$ \Comment{Eq.~\eqref{eq:mg_p_asr}}
        \State $\mathbf{x}_t \gets \frac{1}{\sqrt{\alpha_t}}\left( \mathbf{x}_t^{\textrm{in}}- \frac{1-\alpha_t}{\sqrt{1-\bar{\alpha}_t}} \hat{\boldsymbol{\epsilon}} \right)$
    \Else
        \State $\mathbf{x}_t \gets \frac{1}{\sqrt{\alpha_t}}\left( \mathbf{x}_t^{\textrm{in}}  - \frac{1-\alpha_t}{\sqrt{1-\bar{\alpha}_t}} \boldsymbol{\epsilon}_{\text{cfg}} \right)$
    \EndIf
    \If{$t > 0$}
        \State $\mathbf{x}_t \gets \mathbf{x}_t + \sigma_t \mathbf{z}$ \Comment{Add noise}
    \EndIf
\EndFor
\State $\mathbf{x}_{t=0} \gets \mathbf{x}_\mathrm{m} + \mathbf{x}_{t=0} \odot (1 - \mathbf{m})$ \Comment{Eq.~\eqref{eq:final_Step_inference}}

\State \textbf{Output:} Reconstructed $\mathbf{x}_{t=0}$
\end{algorithmic}
\label{algorithm_difusion_guidance}
\end{algorithm}

\subsection{Models, Architectures, and Training Phase}
This section outlines the core components of our system, detailing the architectural choices and training procedures. We describe backbone options for the conditional DDPM module, present two guidance approaches, and explain the vocoder used for waveform reconstruction.

At the outset, we examined two alternative architectures for the conditional DDPM model used to estimate the diffusion noise, based on either U-Net or \ac{DiT} architectures.
To train the DDPM model, it is necessary to employ a model that estimates the noise from the noisy mel-spectrogram, as described by
\begin{equation}
    \mathbf{x}_t = \sqrt{\bar{\alpha}_t} \mathbf{x}_0 + \sqrt{1 - \bar{\alpha}_t} \mathbf{\boldsymbol{\epsilon}},
    \label{eq:noise_eq}
\end{equation}
given the vectorized clean mel-spectrogram $\mathbf{x}_0 \in \mathbb{R}^d$ and the corresponding timestep $t$. For both architectures, we initially adopted a training objective based on minimizing the \ac{MSE} loss between the reconstructed and target mel-spectrograms. However, empirical results showed improved performance when optimizing the \ac{MSE} loss between the true noise $\boldsymbol{\epsilon}$ and its estimate, defined as:
\begin{equation}
    \mathcal{L} = \mathbb{E}_{t, \mathbf{x}_0, \boldsymbol{\epsilon}} \left[ \| \boldsymbol{\epsilon} - \boldsymbol{\epsilon}_{\boldsymbol{\theta}}(\mathbf{x}_t,\cdot, t) \|^2 \right].
    \label{eq:loss_function}
\end{equation}
where the placeholder is set to $\mathbf{x}_\mathrm{m}$ with probability 0.8, and to $\varnothing$ otherwise.
This finding aligns with the observations reported in \cite{ho2020denoising}.
It should be noted that the DDPM receives $\mathbf{x}_t$ as input during training and $\mathbf{x}_t^{\textrm{in}}$ during inference.

\subsubsection{U-net Archticture}
We investigate the U-Net architecture, adopting the design in~\cite{ho2020denoising}. In our work, this U-Net serves as a baseline for comparison with the \ac{DiT}-based approach.
Following \cite{grad-tts}, we utilize half the number of channels relative to the original U-Net and adopt three feature map resolutions rather than four. Our experiments utilize 80-dimensional mel-spectrograms, with the network operating at progressively downsampled resolutions $80 \times L$, $40 \times L/2$, and $20 \times L/4$. In cases where the number of frames $L$ is not a multiple of four, zero-padding is applied to the mel-spectrograms. Furthermore, following \cite{grad-tts}, the timestep $t$ is embedded using sinusoidal positional encoding.

To incorporate speaker-related information, we condition the U-Net model on WavLM \cite{wavlm} representations. WavLM is a self-supervised speech model based on Wav2Vec 2.0 \cite{wav2vec2}, designed to capture phonetic, prosodic, and speaker-related information while demonstrating robustness to noise and overlapping speech. The input to WavLM is the masked speech signal in the time domain, and its output has a temporal resolution of 
$\frac{L}{2}$ frames. To match it to the temporal resolution of the mel-spectrogram, we apply a transposed convolution layer, increasing the temporal dimension by a factor of two, and then concatenate the resulting representation with the noisy mel-spectrogram as an input to the U-Net model. Since WavLM produces multiple intermediate layer outputs, we introduce learnable weights to aggregate these representations. After training, we analyze the learned weights to determine which layers make the most significant contributions to performance. Empirical results indicate that the first three layers have the most substantial impact, and therefore, we restrict the conditioning to these layers.

\subsubsection{{DiT} Archticture}
Our implementation of the \ac{DiT} module is based on the architecture introduced in \cite{f5_tts}, where \ac{DiT} blocks are integrated with zero-initialized \ac{adaLN} to enhance training stability and conditioning effectiveness. Unlike conventional diffusion models that employ a U-Net backbone, the \ac{DiT} model utilizes a transformer-based architecture, which has been demonstrated to enhance scalability and performance in image generation tasks \cite{DiT_model}. This zero-initialization strategy for \ac{adaLN} layers has been demonstrated to improve model convergence and enhance learning dynamics \cite{adaln_zero}. The input to the \ac{DiT} model is the noisy mel-spectrogram concatenated with the masked mel-spectrogram $\mathbf{x}_\mathrm{m}$ along the feature dimension. 
The concatenated input is first projected to the model’s internal feature dimension through a linear layer. A convolutional positional embedding is then computed from the projected features and added to them, allowing the model to capture positional information.
Unlike the U-Net baseline, the \ac{DiT}-based model does not require external speaker embeddings such as WavLM. We attribute this to the increased capacity of the \ac{DiT} model and stronger inductive bias, which enable it to learn speaker-related information directly from the mel-spectrogram input. This eliminates the need for explicit conditioning on pretrained representations, simplifying the architecture and reducing dependency on external models.

\subsubsection{Guidance architecture}
We investigate two approaches for text-based guidance: phoneme-classifier guidance and \ac{ASR}-classifier guidance. In both cases, we employ the model architecture proposed in \cite{asr_classifier}, with modifications to incorporate diffusion timestep embeddings within the conformer blocks as in \cite{yeminilipvoicer}. For phoneme classification, since the phoneme sequence is inherently longer than the text token sequence, we adjust the subsampling layer in the original model by setting its stride to one instead of two. This modification ensures that the temporal dimension remains sufficiently large for phoneme sequences, preventing excessive compression.

\subsubsection{Vocoder architecture}
To reconstruct the time-domain waveform from the predicted mel spectrograms, we employ HiFi-GAN \cite{kong2020hifi} as a vocoder. The vocoder is trained specifically on our dataset using the same mel-spectrogram configuration as the inpainting model, ensuring consistency between training and inference conditions. We also experimented with a more recent vocoder, BigVGAN \cite{lee2023bigvgan}, which has been shown to produce high-fidelity audio. However, our evaluations indicated that both vocoders yielded comparable perceptual quality in the context of the inpainting task. Consequently, we proceed with HiFi-GAN due to its lower computational cost and established stability.

\section{Experimental Study}
\label{exp}
\subsection{Setup}
\label{exp:setup}
To train and evaluate the conditional DDPM model, we utilized the LibriSpeech dataset, a widely adopted corpus for speech recognition research. LibriSpeech comprises approximately 1,000 hours of English audiobook recordings sourced from LibriVox, accompanied by high-quality transcriptions \cite{librispeech}. Specifically, we employed the \textbf{train-clean-100} and \textbf{train-clean-360} subsets for training, and the \textbf{test-clean} subset for evaluation. The evaluation set comprises speakers who are not included in the training set.
All utterances are sampled at 16 kHz. 

The parameters of the \ac{STFT} are set to a window size of 640 samples and a hop size of 160 samples. The magnitude of the STFT was used to extract 80 mel-frequency components spanning the range 20 Hz to 8 kHz. To ensure numerical stability, values were clipped to a minimum of $1e^{-5}$, and logarithmic compression was applied to mitigate dynamic range issues. Finally, the resulting features were linearly mapped to the range $[-1,1]$, based on the global minimum and maximum values across the training set.

For training the DDPM model, masked speech signals were generated by setting specific regions of the signal to zero while ensuring that no masking occurred at the beginning or end of the utterance. A margin of 0.5 seconds was maintained at both the start and end of each sample. The minimum separation between consecutive masked regions was set to 0.3 seconds, while the duration of each masked segment was randomly sampled in the range of 0.3 to 0.65 seconds. The number of masked regions was determined randomly, with a minimum number of regions given by:
\begin{equation}
\max\left( 1, \left\lfloor \frac{N - 2 \times 0.5}{0.3 + 0.65} \right\rfloor \right) = \max\left( 1, \left\lfloor \frac{N - 1}{0.95} \right\rfloor \right),
\end{equation}
where $N$ denotes the duration of the sample in seconds.

For the \ac{DiT} model, as the masked mel-spectrogram serves as the conditional input, the loss function defined in~\eqref{eq:loss_function} is computed only over the masked regions. In contrast to the DiT model, where the masked mel-spectrogram is fused with the noisy mel-spectrogram before being fed into the model, the U-Net model first processes the masked signal in the time domain through the WavLM encoder and subsequently fuses it with the noisy mel-spectrogram. Due to this difference in integration strategy, the loss function for the U-Net is not biased toward the masked regions. This contrasts with the DiT setup, where, based on empirical observations, we assigned 80\% of the total loss to the masked regions and 20\% to the unmasked regions.

The models were trained for 5 million mini-batches, with each batch containing 32 speech signals. We utilized the Adam optimizer with a fixed learning rate of $2 \times 10^{-4}$, without any learning rate scheduling. For classifier-free guidance, we followed the methodology proposed in \cite{classifier_free}, setting the dropout probability for conditioning to 0.2. When conditioning was omitted as part of the classifier-free guidance strategy, the null input was set to all zeros. For the U-Net model, this corresponds to a zeroed masked speech signal, while for the DiT model, it corresponds to a zeroed masked mel-spectrogram.

For evaluation, we conducted two separate settings to assess the performance of the inpainting model under different conditions.
In the first setting, where the corresponding text is provided, we randomly selected 50 samples from the \textbf{test-clean} subset, ensuring that each sample had a duration exceeding 7 seconds. Masking was applied with varying gap durations: 1 second, 0.5 seconds, and 0.25 seconds, with each masked segment separated by a 1.5-second unmasked region. These configurations are referred to as large, medium, and small masking gaps, respectively. To preserve contextual information at the sample boundaries, a margin of 1.5 seconds was maintained at both the beginning and end of each utterance.
It is important to highlight that these gap durations differ from those used during training to effectively evaluate the model’s ability to generalize to new masking conditions.

In the second setting, where the model operates without access to the given text, we constructed masked signals using shorter gap durations of 0.1, 0.2, and 0.3 seconds, applied at intervals of 0.25, 0.5, and 0.75 seconds, respectively. These shorter masks were designed to avoid fully obscuring key words, which are terms whose absence introduces multiple plausible completions, thereby preserving the ability of the ASR+LM system to infer the missing content. In contrast, longer masks may significantly hinder reconstruction by removing essential semantic information. Thus, the chosen durations strike a balance between creating a meaningful challenge and maintaining recoverability from the surrounding acoustic context.
\subsection{Objective Metics}
To comprehensively evaluate the quality of the inpainted speech signals, we employed several objective metrics.

First, we applied the recently proposed SpeechBLEU \cite{saeki2024speechbertscore} measure. Inspired by the traditional \ac{BLEU} metric \cite{papineni2002bleu}, SpeechBLEU adapts the n-gram precision scoring to discrete speech tokens, thereby enabling the evaluation of generated speech against reference utterances. In our experiments, we utilized the Hubert-base model with a sampling rate of $16$~kHz, extracted features from the 11th layer, and employed a vocabulary size of 200. We computed 5-gram statistics and removed repeated tokens prior to scoring.

While we initially considered using the non-intrusive \ac{MOS}, we found it unsuitable for our task. In cases with large masked regions (e.g., 1-second gaps), the non-intrusive \ac{MOS} yields unrealistically high scores, as it lacks access to the ground-truth reference and therefore cannot detect missing or incorrect content within the masked segments. To address this limitation, we adopted the Distance \ac{MOS}, also referred to as \ac{NMR} in \cite{scoreq2024}, which predicts the perceptual quality by estimating the distance between clean and generated speech samples. As its name implies, the score is lower-bounded by zero and unbounded above, with lower scores reflecting better perceptual quality.

In addition to perceptual metrics, we evaluated the transcription accuracy of the inpainted speech using \ac{ASR}. Specifically, we used a pre-trained Whisper model \cite{whisprt_asr} and reported the resulting \ac{WER}. It is important to emphasize that the \ac{ASR} model used for evaluation differs from the one employed during the guidance process. 

To assess the acoustic fidelity and prosodic consistency of the generated speech, we further calculated two widely used synthesis metrics: the \ac{MCD} and the \ac{LogF0-MSE}. \ac{MCD} measures the spectral distance between generated and reference speech by comparing mel-cepstral coefficients, with \ac{DTW} applied to align sequences of differing lengths. Similarly, \ac{LogF0-MSE} evaluates prosodic accuracy by computing the mean squared error between the aligned log-F0 contours of the generated and reference speech.

\subsection{Baseline Method}
As a baseline method, we computed the evaluation metrics on speech samples generated by a \ac{TTS} model conditioned only on the masked transcripts. Specifically, we trained the StyleSpeech model \cite{lou2024stylespeech} with our mel-spectrogram transform settings. This model generates a speech signal given a desired phoneme sequence and a reference speaker signal. In our case, the reference speech signal is the masked input. We chose this \ac{TTS} model as it provides good results in generating speech with characteristics close to those of the reference signal.

\subsection{Experimental Results}
\label{sec:experiments}
This section presents a comprehensive evaluation of the proposed inpainting models under a range of conditions. We begin by optimizing key weighting parameters, followed by quantitative evaluations conducted both with and without access to the transcript. This structured analysis reveals the models’ strengths and limitations, and examines the influence of guidance and gap duration on inpainting performance.
\subsubsection{Optimization of Weighting Parameters}
The weighting parameters \(w_1\) and \(w_2\) play a crucial role in balancing the contributions of the diffusion prior and classifier guidance during inpainting. To determine the optimal pair of weighting parameters $w_1$ and $w_2$, we performed an exhaustive evaluation over all combinations of $w_1 \in \{-1, 0.5, 0.8, 1, 2\}$ and $w_2 \in \{0.5, 0.7, 0.8, 0.9, 1, 1.2, 1.5, 2, 3\}$ using the validation set described in Section~\ref{exp:setup}. This search was conducted separately for each model variant (\ac{DiT} and U-Net) and for each type of guidance (\ac{ASR} and Phoneme). The combination that yielded the best performance on the validation set was selected as the optimal configuration. These optimal weights were then fixed and directly applied to the unseen test samples, without further tuning.

Analyzing the influence of $w_1$, which controls the guidance-free diffusion prior, reveals that decreasing $w_1$ leads to a deterioration in all evaluation metrics. This occurs because a lower $w_1$ reduces the influence of the original diffusion model, leading to instability and a loss of naturalness in the generated samples.
On the other hand, when $w_2$ is set too high, the classifier guidance dominates the generation process, potentially causing overfitting to the guidance (i.e., the target phoneme or text sequence) and introducing unnatural artifacts or distortions in the output.

Given the complexity of the speech inpainting task, no single metric is sufficient to capture the system’s performance comprehensively. Each metric emphasizes different aspects of quality; for instance, a low \ac{WER} may still correspond to speech with unnatural prosody. Consequently, we jointly considered multiple metrics when selecting the weighting parameters for the guidance. Specifically, in optimizing $w_1/w_2$, we prioritized achieving the best performance with respect to both \ac{WER} and SpeechBLEU. When different weight combinations yielded similar performance on these primary metrics, we also utilized the Distance \ac{MOS} score as a secondary selection criterion.

\subsubsection{Quantitative and Qualitative Evaluation with Given text}
In this section, we evaluate the performance of our proposed inpainting model when the transcript is available during inference, focusing on both quantitative metrics and qualitative observations.
The results are presented in Table~\ref{table:results_metric}. First, it can be observed that the guidance improves performance across all settings. Second, for all model configurations, as the gap size increases, the evaluation metrics degrade, as expected.

For both models, U-Net and \ac{DiT}, incorporating phoneme guidance consistently leads to improved performance across all metrics. This improvement is attributed to the fact that phoneme guidance enforces a deeper and more fundamental constraint during the generation process. While \ac{ASR} guidance may predict words that sound similar to the target but are incorrect, phoneme guidance directly steers the reverse process at the phoneme level, enforcing a stricter adherence to the correct pronunciation.

Another observation is that the \ac{DiT} model consistently outperforms the U-Net model across all metrics and gap durations when using phoneme guidance. This can be attributed to the transformer-based architecture of \ac{DiT}, which is more effective in capturing long-range dependencies in the speech representations.

Finally, from the no-guidance condition, we observe that the prosody of the generated speech is relatively preserved, as indicated by the \ac{MCD} and \ac{LogF0-MSE} metrics.
Audio samples generated by our inpainting model are available on our demo page\footnote{\href{https://mordehaym.github.io/PGDI/}{https://mordehaym.github.io/PGDI/}}.

\begin{figure*}[htbp]
    \centering
      \includegraphics[width=0.99\textwidth]{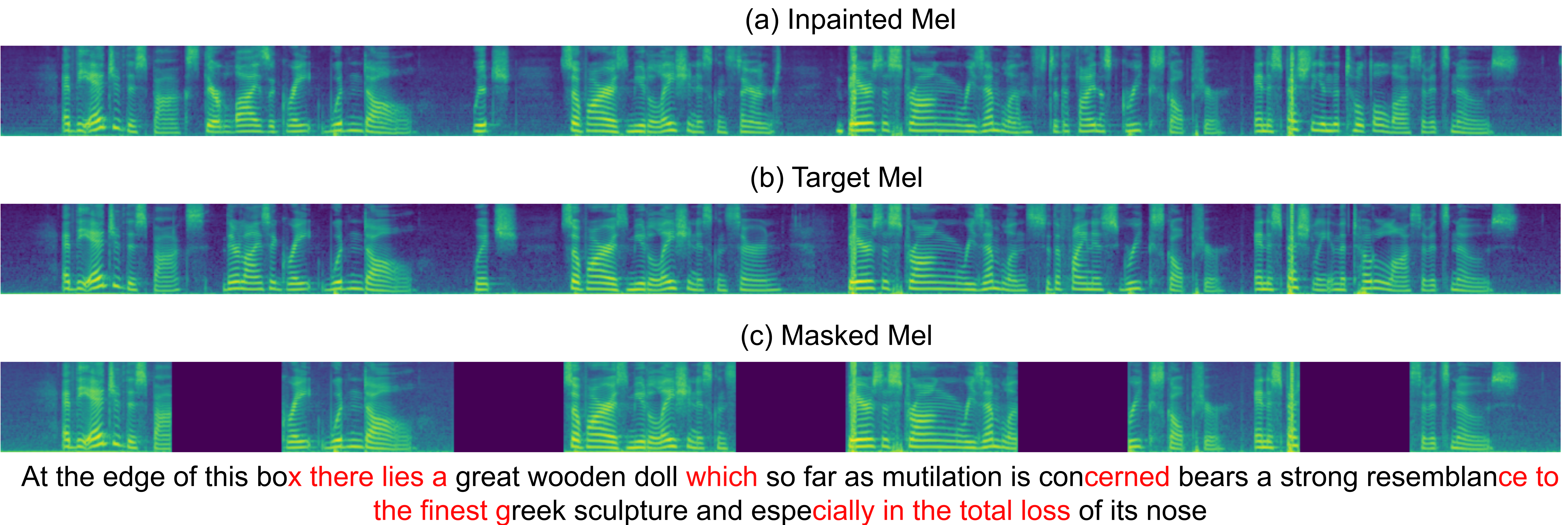}
      \setlength{\belowcaptionskip}{0pt}
      \caption{An example from our evaluation dataset, along with the inpainted result. In this case, the masked segment is 1 second long. The red regions in the transcription correspond to the masked speech segments. The text is given in this example.}
      \label{figs:masked_inpainted_text_sample=244}
\end{figure*}

To further qualitatively demonstrate the effectiveness of the proposed inpainting model, Fig.~\ref{figs:masked_inpainted_text_sample=244} presents a representative example of (a) the inpainted mel-spectrogram, (b) the target spectrogram, and (c) the corresponding masked mel spectrogram, along with the associated transcript. 
The masked regions correspond to the red-colored words in the transcript, highlighting the segments removed during inference. As shown, the inpainted spectrogram closely resembles the target in both structure and acoustic continuity, particularly within the previously missing regions. This visual similarity supports our quantitative findings, illustrating the model’s ability to generate semantically and prosodically coherent content conditioned on phoneme guidance.

\begin{table*}[htbp]
 \renewcommand{\arraystretch}{0.9}
 \setlength{\tabcolsep}{4pt}
 \resizebox{0.99\columnwidth}{!}{
\begin{tabular}{@{}ccccccccc@{}}
\toprule
\multirow{2}{*}{\centering \textbf{Model}} & \multirow{2}{*}{\centering \textbf{Gap}} & \multirow{2}{*}{\centering \textbf{Guidance}} & \multirow{2}{*}{\centering \textbf{w1/w2}} & \multicolumn{5}{c}{\textbf{Metrics}} \\
\cmidrule(l){5-9}
& & & & \textbf{\ac{WER}[\%]} $\downarrow$ & \textbf{SpeechBLEU} $\uparrow$ & \textbf{Distance MOS} $\downarrow$ & \textbf{\ac{MCD}} $\downarrow$ & \textbf{\ac{LogF0-MSE}} $\downarrow$ \\
\midrule
\multirow{3}{*}{\centering Unprocessed} 
 & $0.25$ & --         & -- &    5.1   &  0.59     &    0.6   &    3.08   &   0.22    \\
\cmidrule(l){2-9}
 & $0.5$ & --         & -- &    17.36   &  0.52     &   0.56    &    4.24   &   0.23    \\
\cmidrule(l){2-9}
 & $1$   & --         & -- &    32.18   &     0.42  &   0.5    &   5.2    &    0.25   \\
\midrule
\multirow{3}{*}{\centering \ac{TTS}} 
 & $0.25$ & --         & -- &    \underline{\textbf{0}}   &  0.3     &    0.37   &    5.7   &   0.25    \\
\cmidrule(l){2-9}
 & $0.5$ & --         & -- &    \underline{\textbf{0}}   &  0.32     &   0.36    &    5.6   &   0.25    \\
\cmidrule(l){2-9}
 & $1$   & --         & -- &    \underline{\textbf{0}}   &     0.31  &   0.36    &   5.6    &    0.25   \\
\midrule
\multirow{9}{*}{\centering U-Net} 
 & \multirow{3}{*}{$0.25$} & ASR         &  1/0.8   &    \underline{\textbf{0}}   &   0.78    &    0.15   &   0.97    &   \underline{0.11}    \\
 &                          & Phoneme     &  1/0.5   &    \underline{\textbf{0}}   & \underline{0.8} & \underline{\textbf{0.1}} & \underline{0.85} & \underline{0.11}   \\
 &                          & No Guidance &  --      &   13.71    &   0.77    &   0.11    &   1.02    &   \underline{0.11}   \\
\cmidrule(l){2-9}
 & \multirow{3}{*}{$0.5$}   & ASR         &  1/0.5   &   6.51     &   0.68    &    0.19   &    1.75   &   \underline{\textbf{0.13}}    \\
 &                          & Phoneme     &  1/0.7   &   \underline{\textbf{0}}   & \underline{0.7} &  0.21   &  \underline{1.6}     &    0.14   \\
 &                          & No Guidance &  --      &   29.7     &   0.67    & \underline{0.18} &    2.16   &   \underline{\textbf{0.13}}   \\
\cmidrule(l){2-9}
 & \multirow{3}{*}{$1$}     & ASR         &  1/0.9   &   9.72     &   0.58    &    0.32   &    2.95   &   \underline{0.17}    \\
 &                          & Phoneme     &  1/0.5   &   \underline{5.48}  &   \underline{0.61}   &    0.29   &   \underline{2.61}    &   \underline{0.17}   \\
 &                          & No Guidance &  --      &   39.14    &   0.57    & \underline{0.22} &   3.4    &   0.18    \\
\midrule
\multirow{9}{*}{\centering {DiT}} 
 & \multirow{3}{*}{$0.25$} & ASR         &  1/0.8   &    \underline{\textbf{0}}   &     0.8  &    0.13   &   0.84    &   \underline{\textbf{0.1}}   \\
 &                          & Phoneme     &  1/0.5   &    \underline{\textbf{0}}   & \underline{\textbf{0.81}} & \underline{\textbf{0.1}} & \underline{\textbf{0.75}} & \underline{\textbf{0.1}}  \\
 &                          & No Guidance &  --      &   15.59    &   0.77   &    0.11   &    0.95   &   \underline{\textbf{0.1}}    \\
\cmidrule(l){2-9}
 & \multirow{3}{*}{$0.5$}  & ASR         &  1/0.5   &   6.8      &   0.7    &    0.18   &   1.58    &   0.14   \\
 &                          & Phoneme     &  1/0.7   &   \underline{\textbf{0}}   & \underline{\textbf{0.71}} &   0.18    &   \underline{\textbf{1.45}}   &   \underline{\textbf{0.13}}   \\
 &                          & No Guidance &  --      &   30.38    &   0.68    & \underline{\textbf{0.16}} &    1.93   &   \underline{\textbf{0.13}}   \\
\cmidrule(l){2-9}
 & \multirow{3}{*}{$1$}    & ASR         &  0.8/1.2 &   8.71     &   0.59    &   0.31    &   2.86    &   0.17    \\
 &                          & Phoneme     &  1/0.5   &   \underline{5.34}  & \underline{\textbf{0.62}} &   0.27    &   \underline{\textbf{2.48}}   &   \underline{\textbf{0.15}}   \\
 &                          & No Guidance &  --      &   40.37    &   0.56    & \underline{\textbf{0.21}} &    3.14   &   0.18   \\
\bottomrule
\end{tabular}
}
\caption{Evaluation results of Unprocessed signal, baseline \ac{TTS}, U-Net, and \ac{DiT} under different gap and guidance conditions across multiple metrics, when the transcript is available at inference time.
\textbf{Bold} indicates the best result for a given gap duration across all models and guidance types. 
\underline{Underline} indicates the best result for a given gap duration within a specific model across guidance types.}
\label{table:results_metric}
\end{table*}

\subsubsection{Quantitative and Qualitative Evaluation without Given Text}
\begin{figure}[htbp]
    \centering
    
    \begin{tabular}{@{}c@{\hspace{0.02\linewidth}}c@{\hspace{0.02\linewidth}}c@{\hspace{0.02\linewidth}}c@{}}
        & \textsc{\footnotesize 0.1s gap} & \textsc{\footnotesize 0.2s gap} & \textsc{\footnotesize 0.3s gap} \\[0.5em]
        
        \rotatebox{90}{\hspace{1.15cm} \textsc{\footnotesize WER}} &
        \begin{tabular}[b]{@{}c@{}}
            \includegraphics[width=0.32\linewidth]{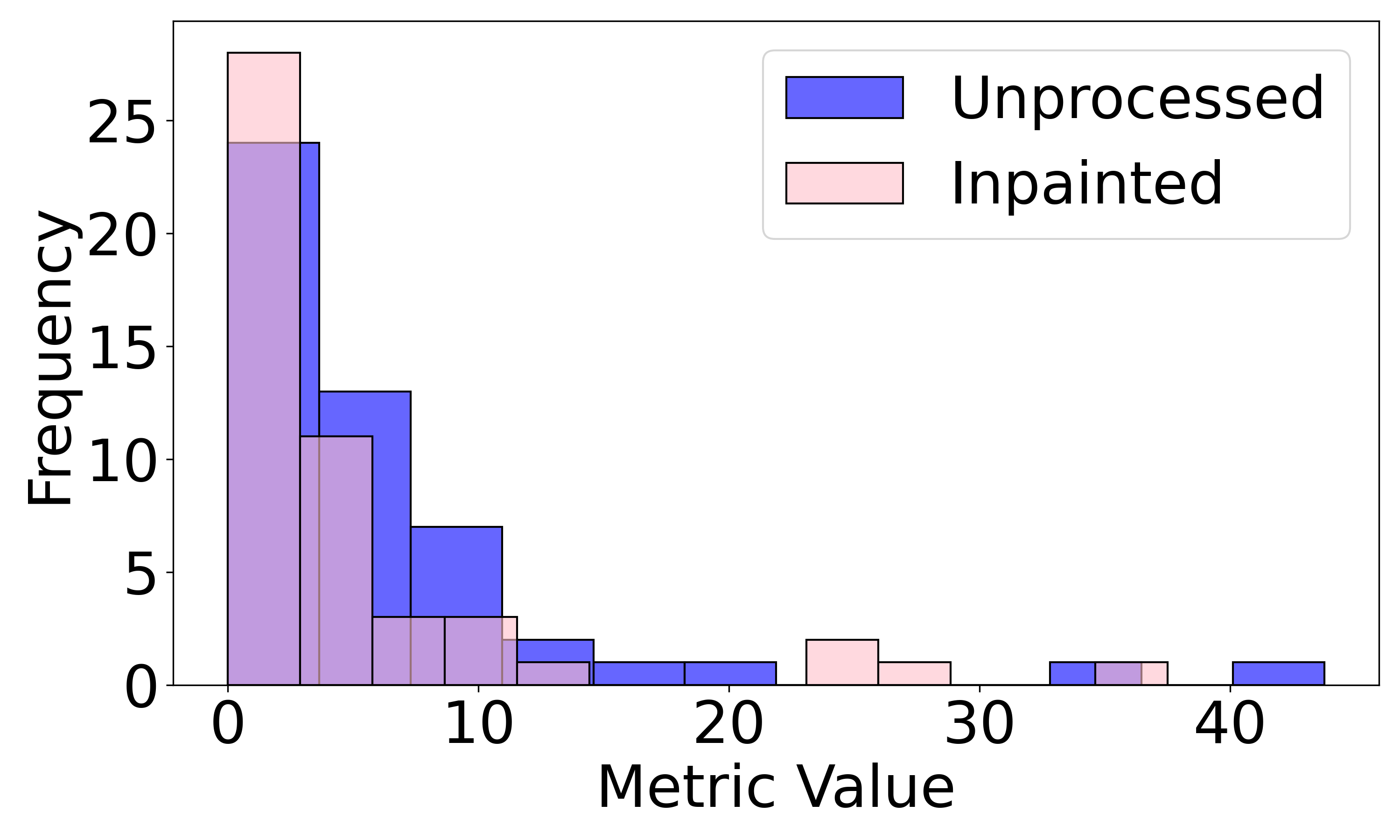} \\
            \tiny WER = 3.94 $\xrightarrow{}$ 2.47
        \end{tabular} &
        \begin{tabular}[b]{@{}c@{}}
            \includegraphics[width=0.32\linewidth]{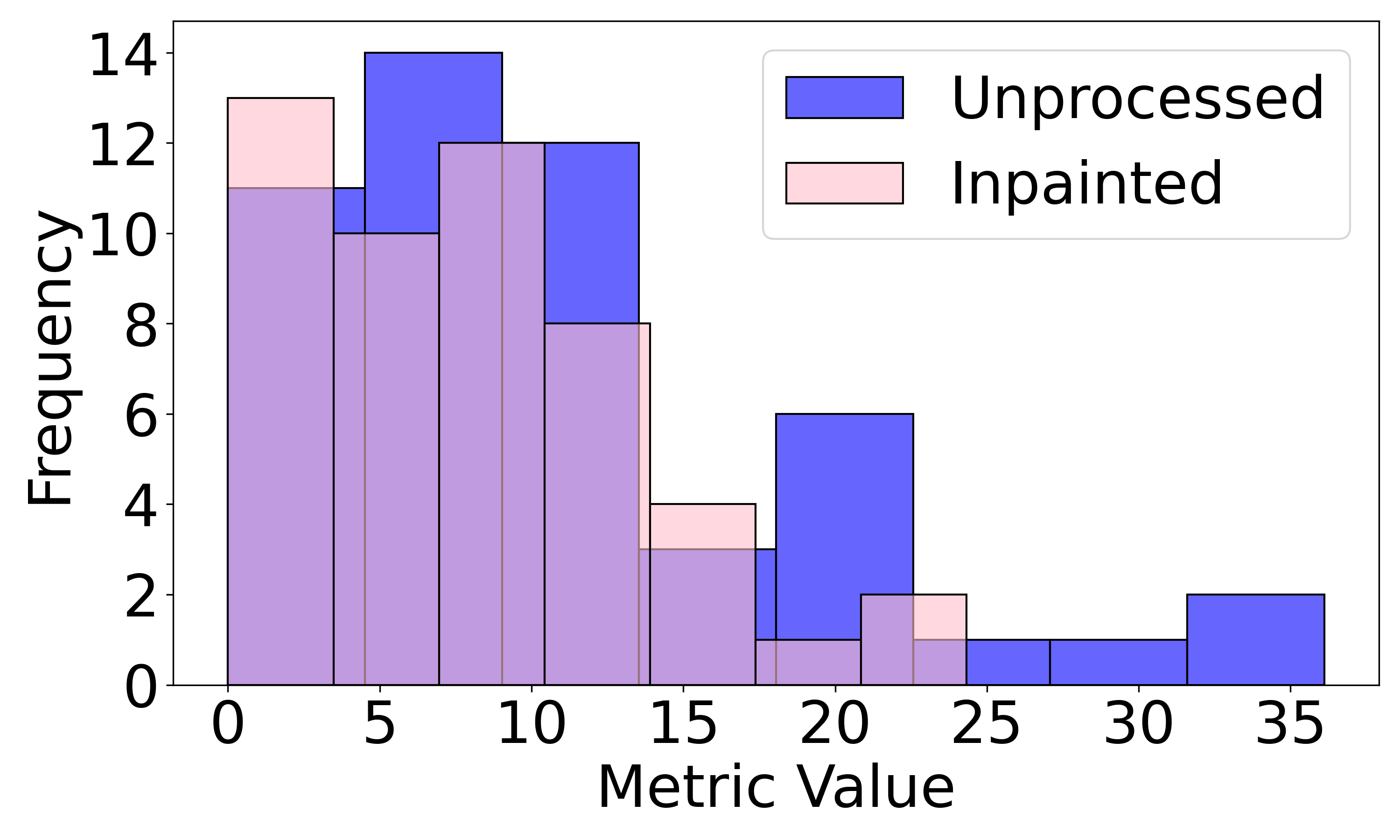} \\
            \tiny WER = 8.2 $\xrightarrow{}$ 7.41
        \end{tabular} &
        \begin{tabular}[b]{@{}c@{}}
            \includegraphics[width=0.32\linewidth]{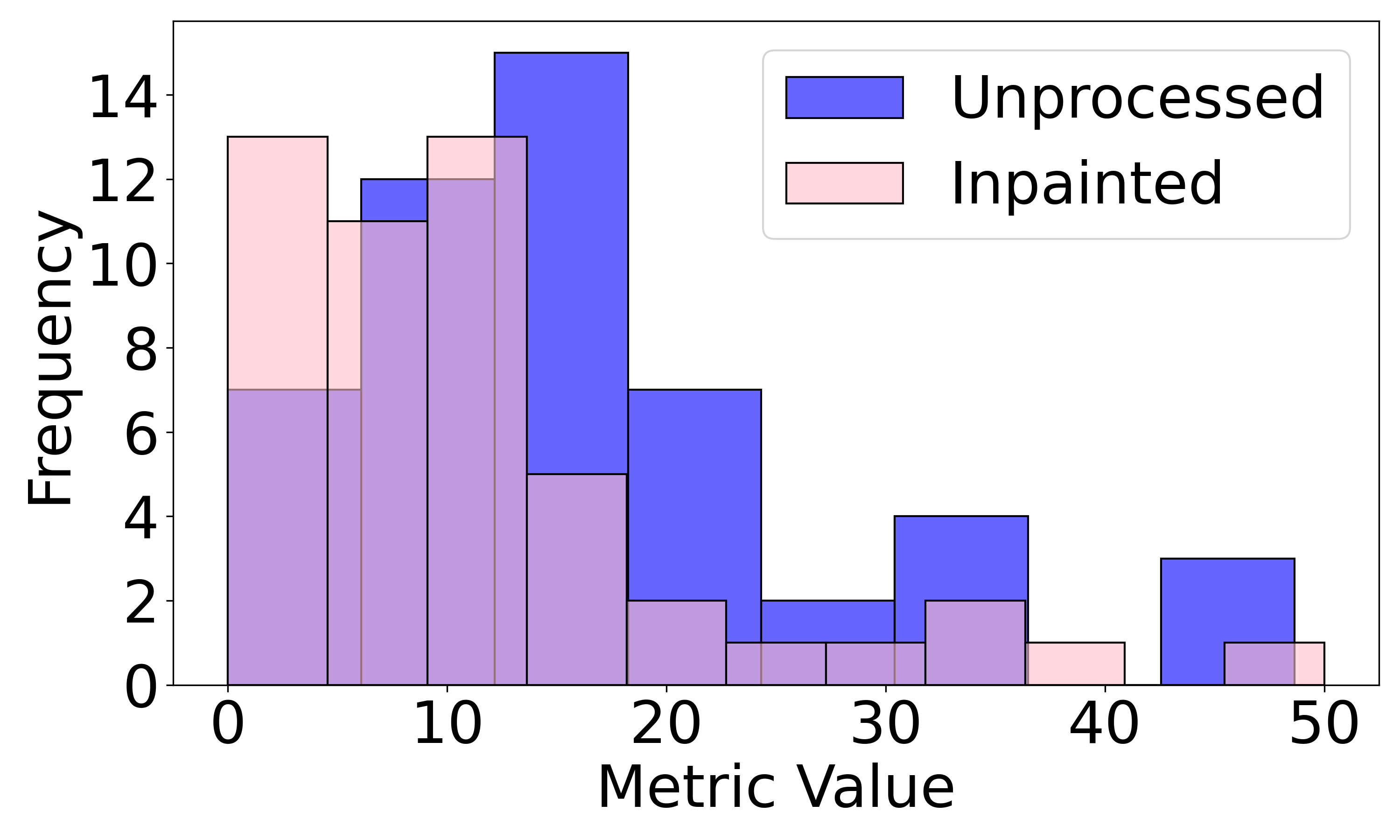} \\
            \tiny WER = 13.71 $\xrightarrow{}$ 9.45
        \end{tabular} \\[0.5em]
        
        \rotatebox{90}{\hspace{0.32cm} \textsc{\footnotesize SpeechBLEU}} &
        \begin{tabular}[b]{@{}c@{}}
            \includegraphics[width=0.32\linewidth]{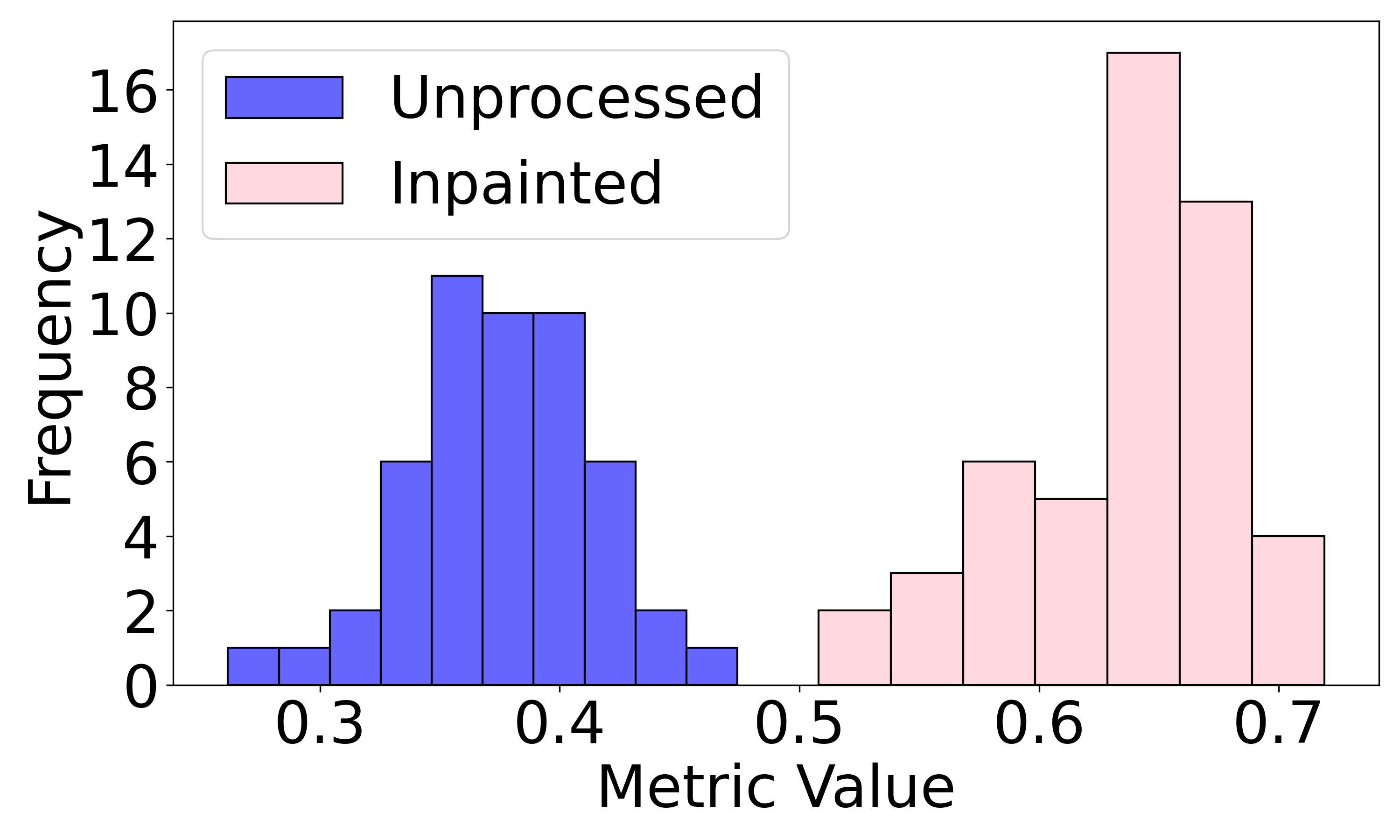} \\
            \tiny SpeechBLEU = 0.38 $\xrightarrow{}$ 0.65
        \end{tabular} &
        \begin{tabular}[b]{@{}c@{}}
            \includegraphics[width=0.32\linewidth]{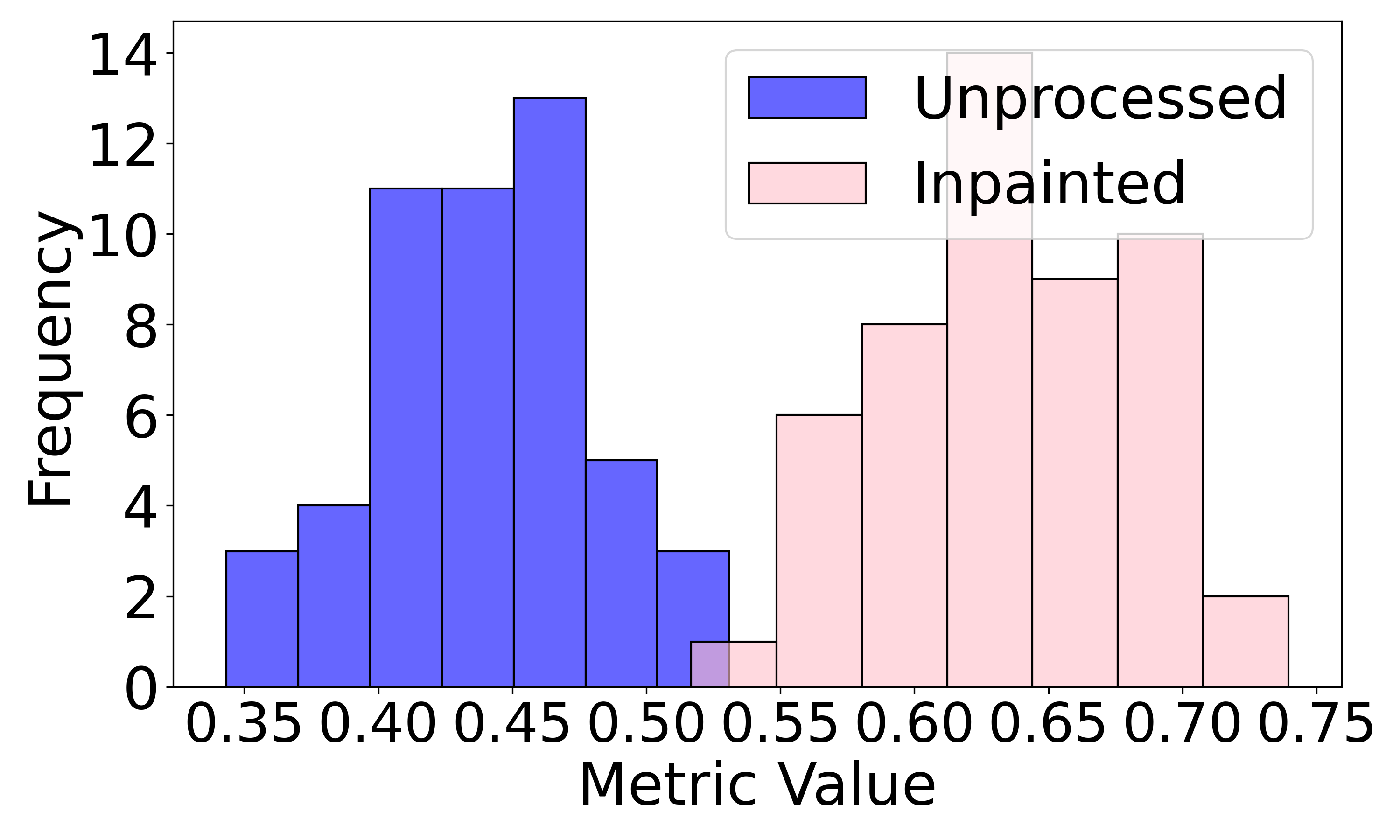} \\
            \tiny SpeechBLEU = 0.44 $\xrightarrow{}$ 0.63
        \end{tabular} &
        \begin{tabular}[b]{@{}c@{}}
            \includegraphics[width=0.32\linewidth]{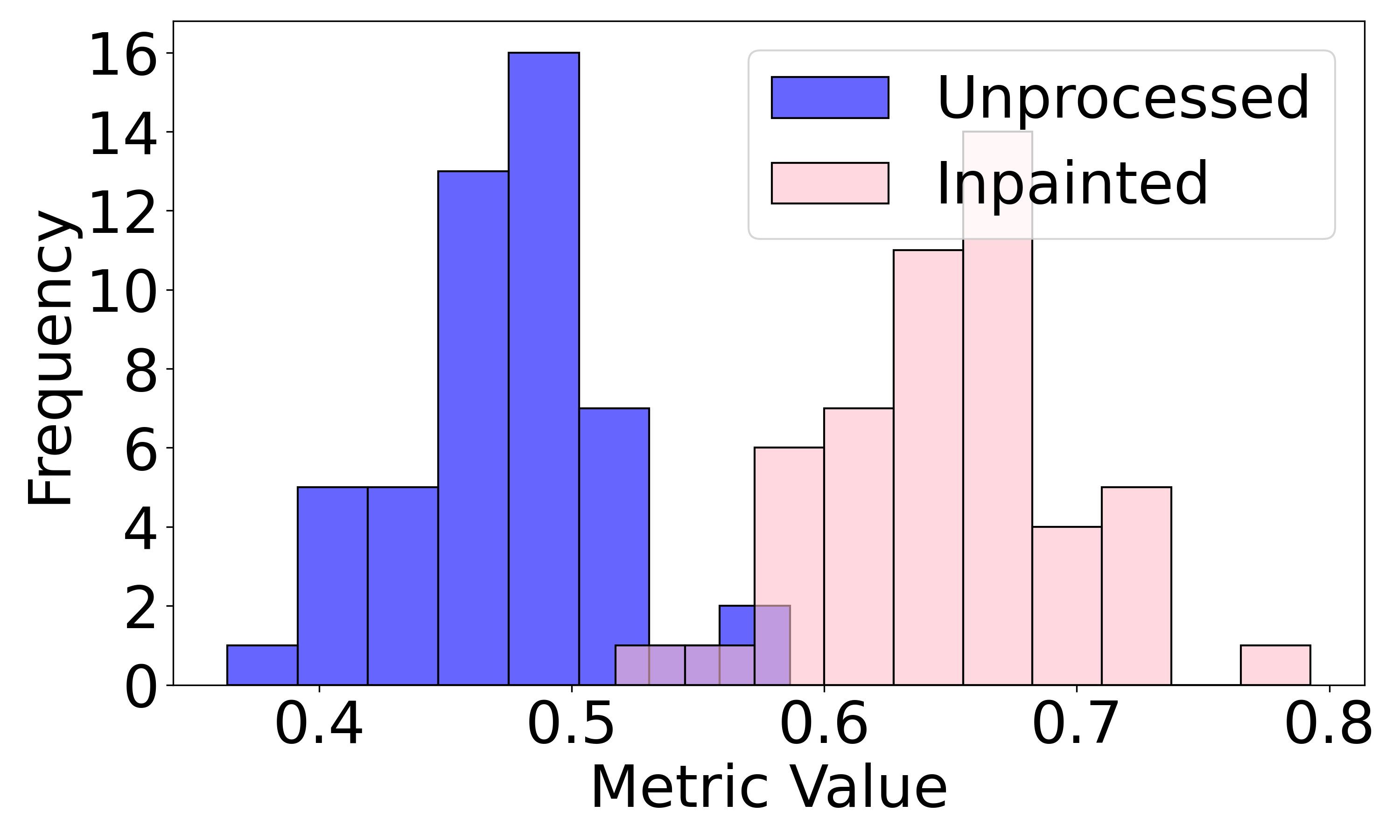} \\
            \tiny SpeechBLEU = 0.48 $\xrightarrow{}$ 0.65
        \end{tabular} \\[0.5em]
        
        \rotatebox{90}{\hspace{0.37cm}\textsc{\footnotesize Distance MOS}} &
        \begin{tabular}[b]{@{}c@{}}
            \includegraphics[width=0.32\linewidth]{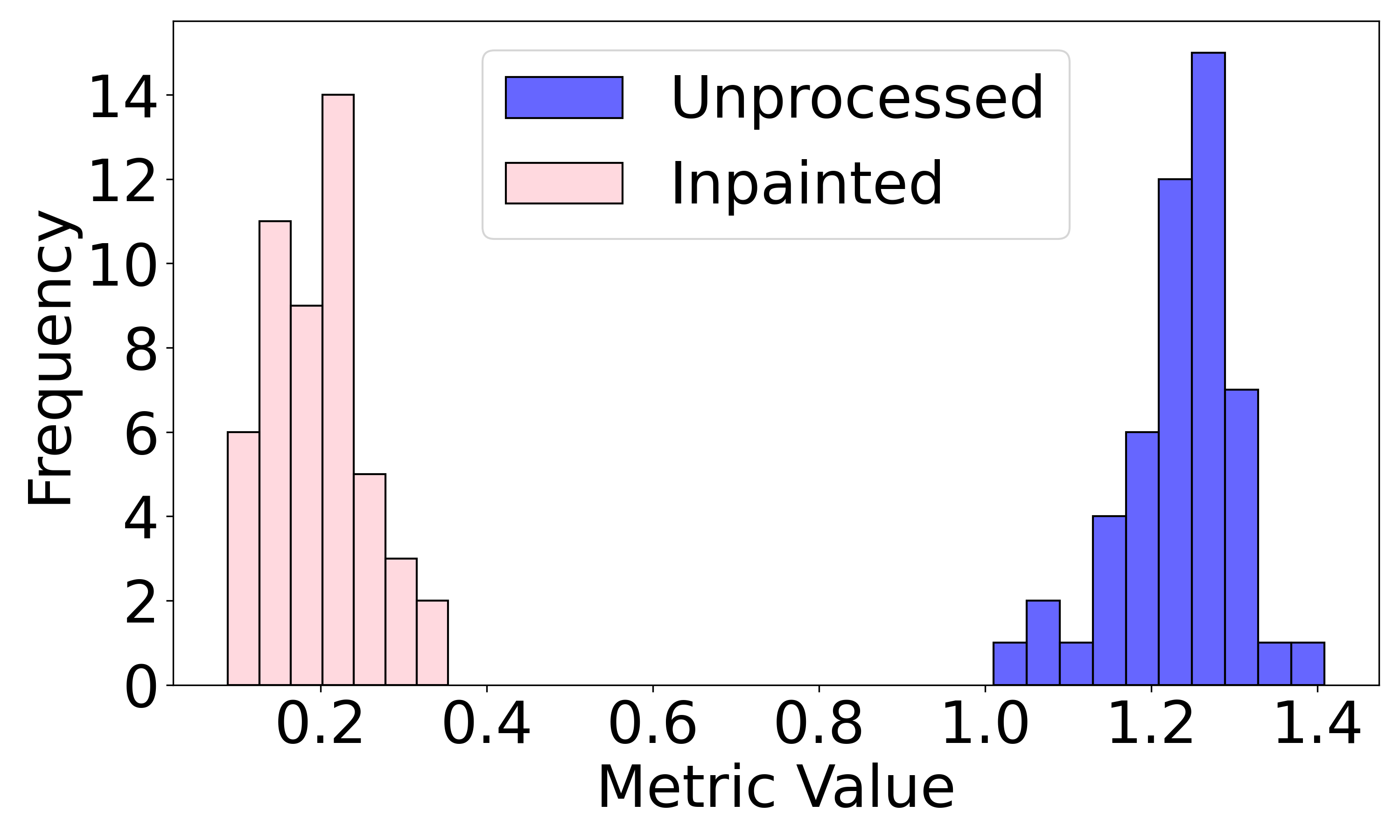} \\
            \tiny Distance MOS = 1.24 $\xrightarrow{}$ 0.19
        \end{tabular} &
        \begin{tabular}[b]{@{}c@{}}
            \includegraphics[width=0.32\linewidth]{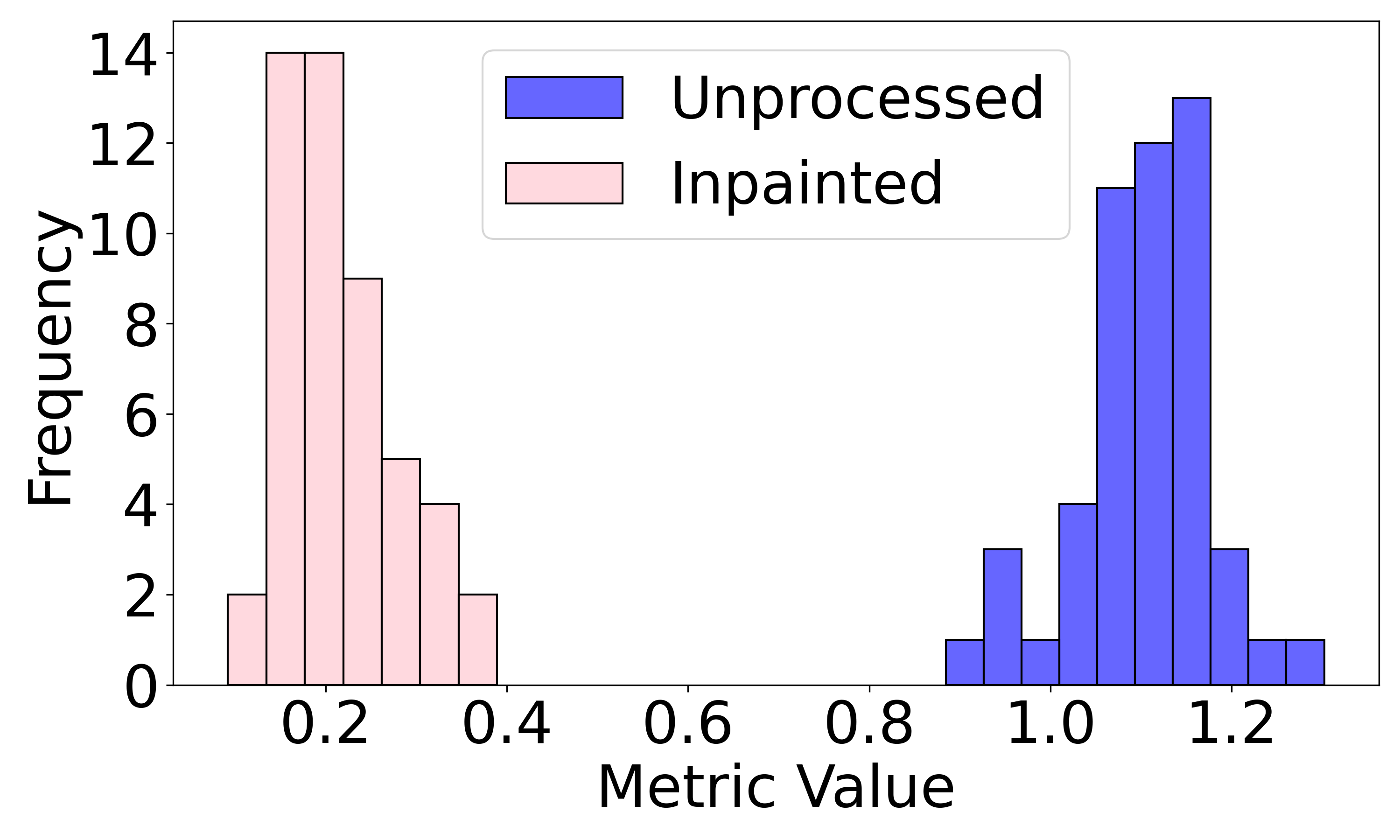} \\
            \tiny Distance MOS = 1.12 $\xrightarrow{}$ 0.19
        \end{tabular} &
        \begin{tabular}[b]{@{}c@{}}
            \includegraphics[width=0.32\linewidth]{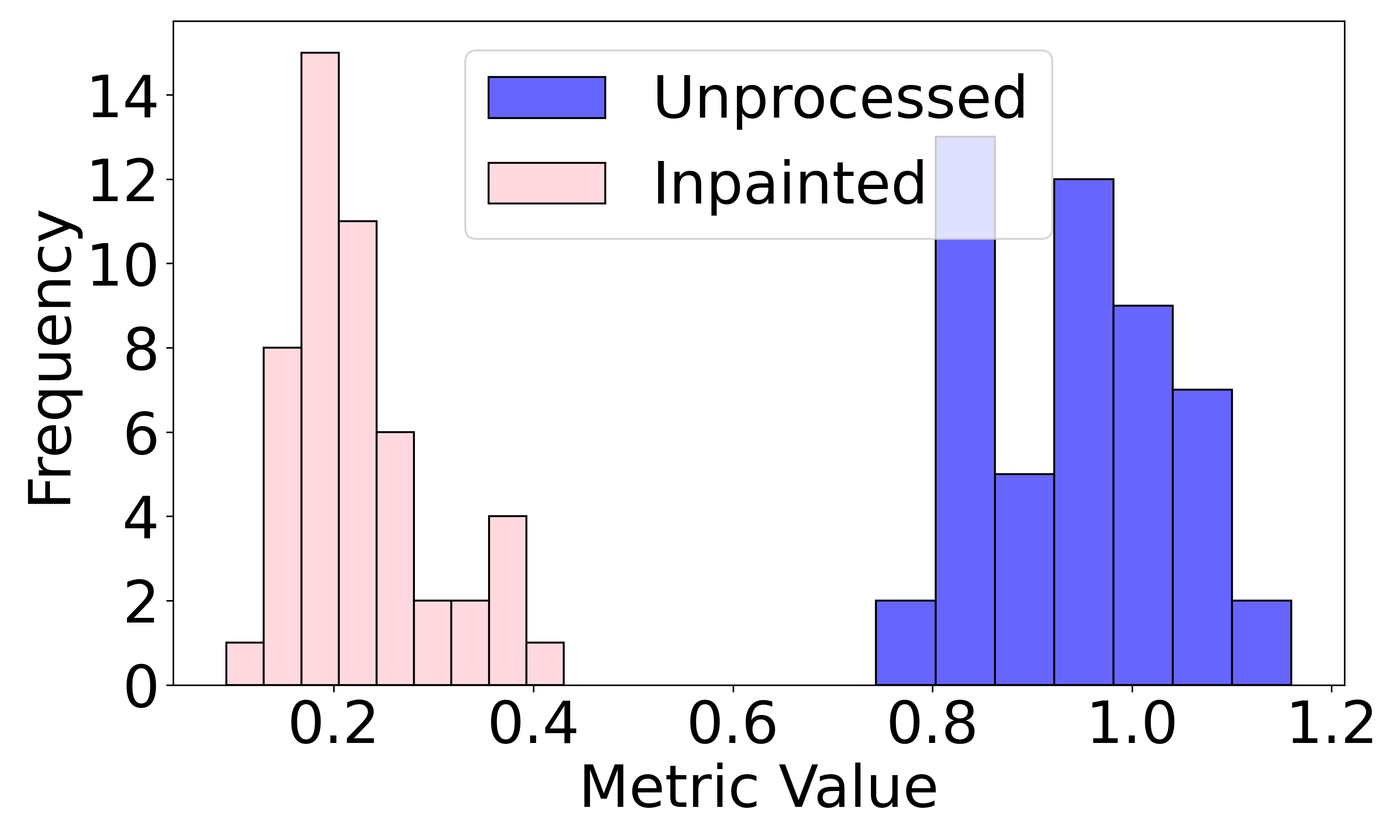} \\
            \tiny Distance MOS = 0.95 $\xrightarrow{}$ 0.21
        \end{tabular} \\[0.5em]
        
        \rotatebox{90}{\hspace{1.6cm}\textsc{\footnotesize MCD}} &
        \begin{tabular}[b]{@{}c@{}}
            \includegraphics[width=0.32\linewidth]{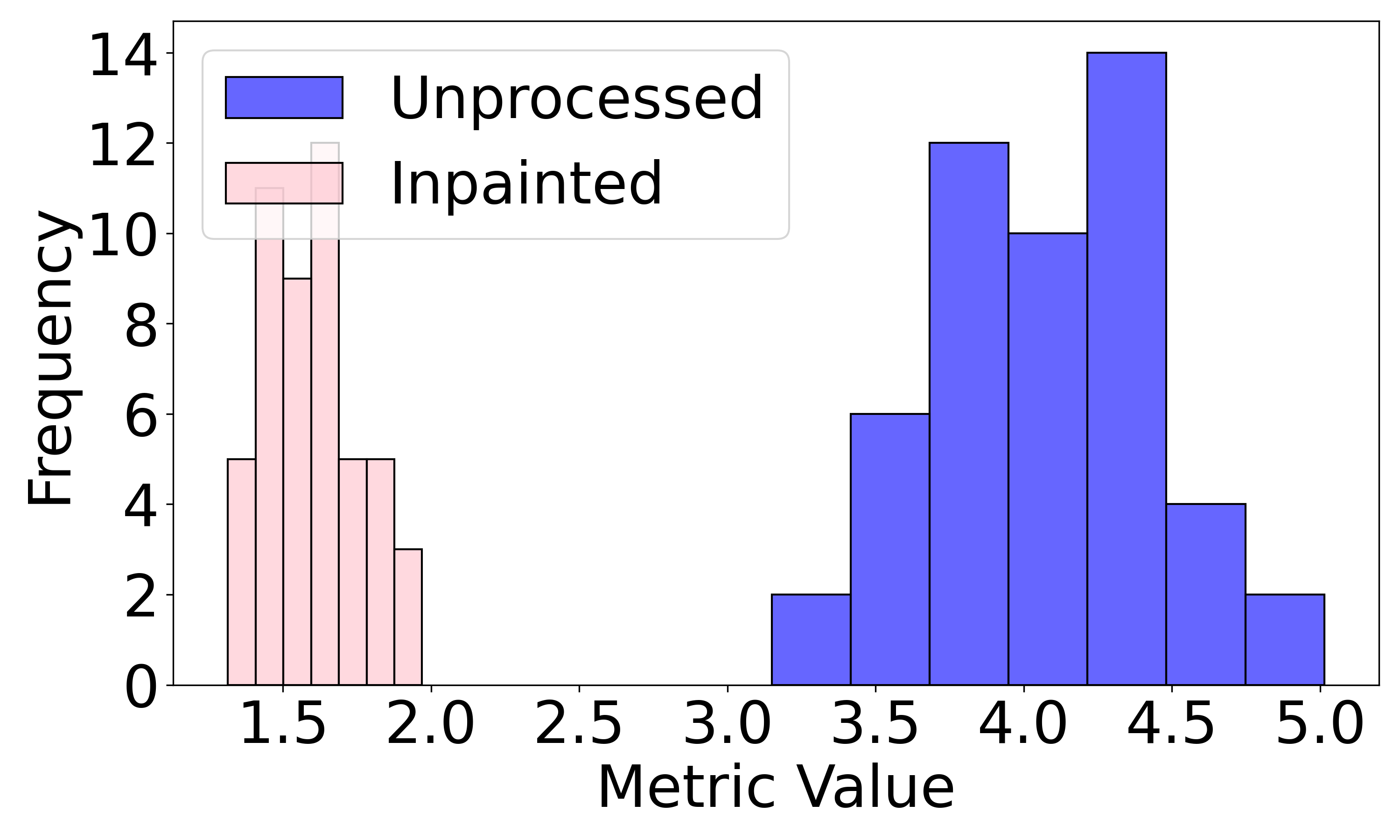} \\
            \tiny MCD = 4.14 $\xrightarrow{}$ 1.60
        \end{tabular} &
        \begin{tabular}[b]{@{}c@{}}
            \includegraphics[width=0.32\linewidth]{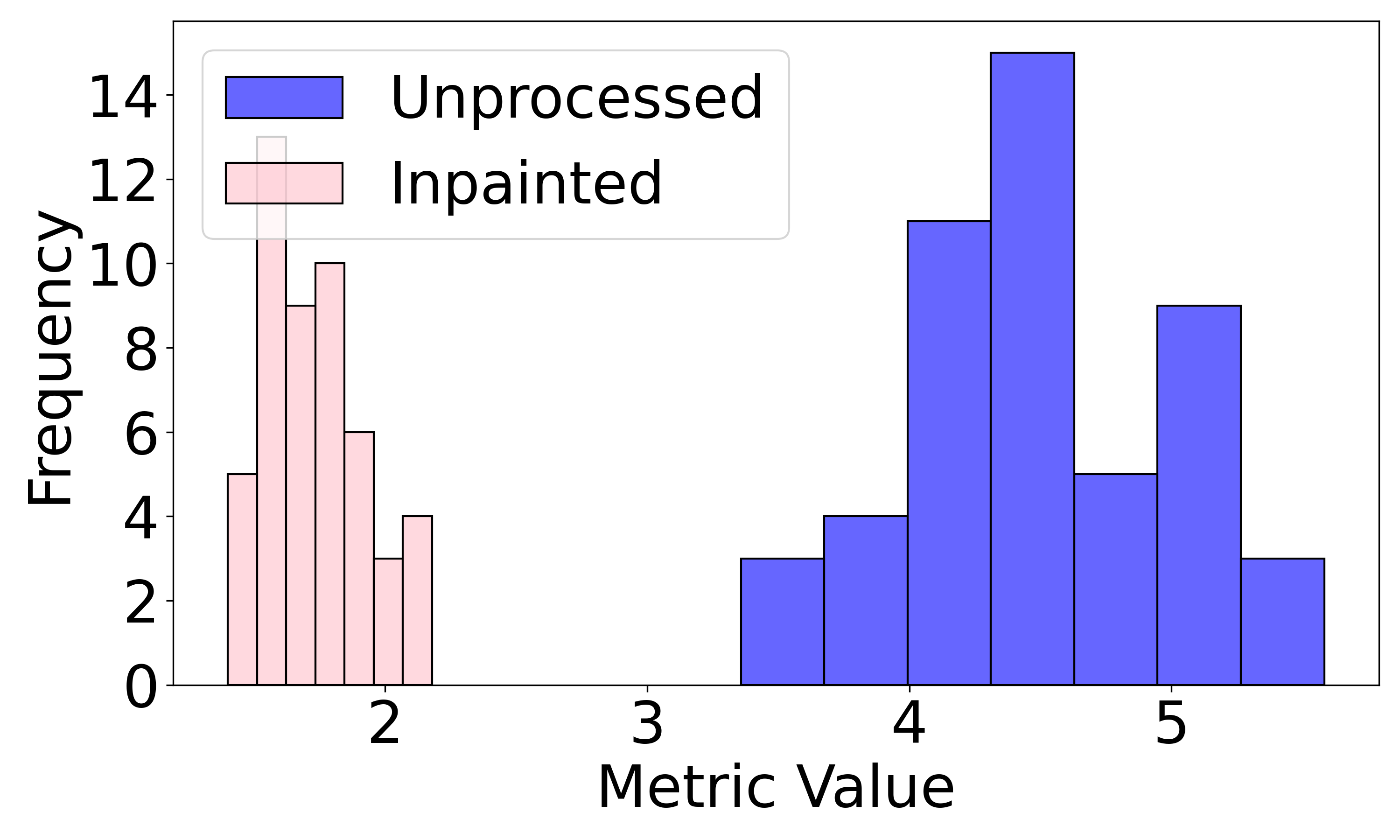} \\
            \tiny MCD = 4.49 $\xrightarrow{}$ 1.70
        \end{tabular} &
        \begin{tabular}[b]{@{}c@{}}
            \includegraphics[width=0.32\linewidth]{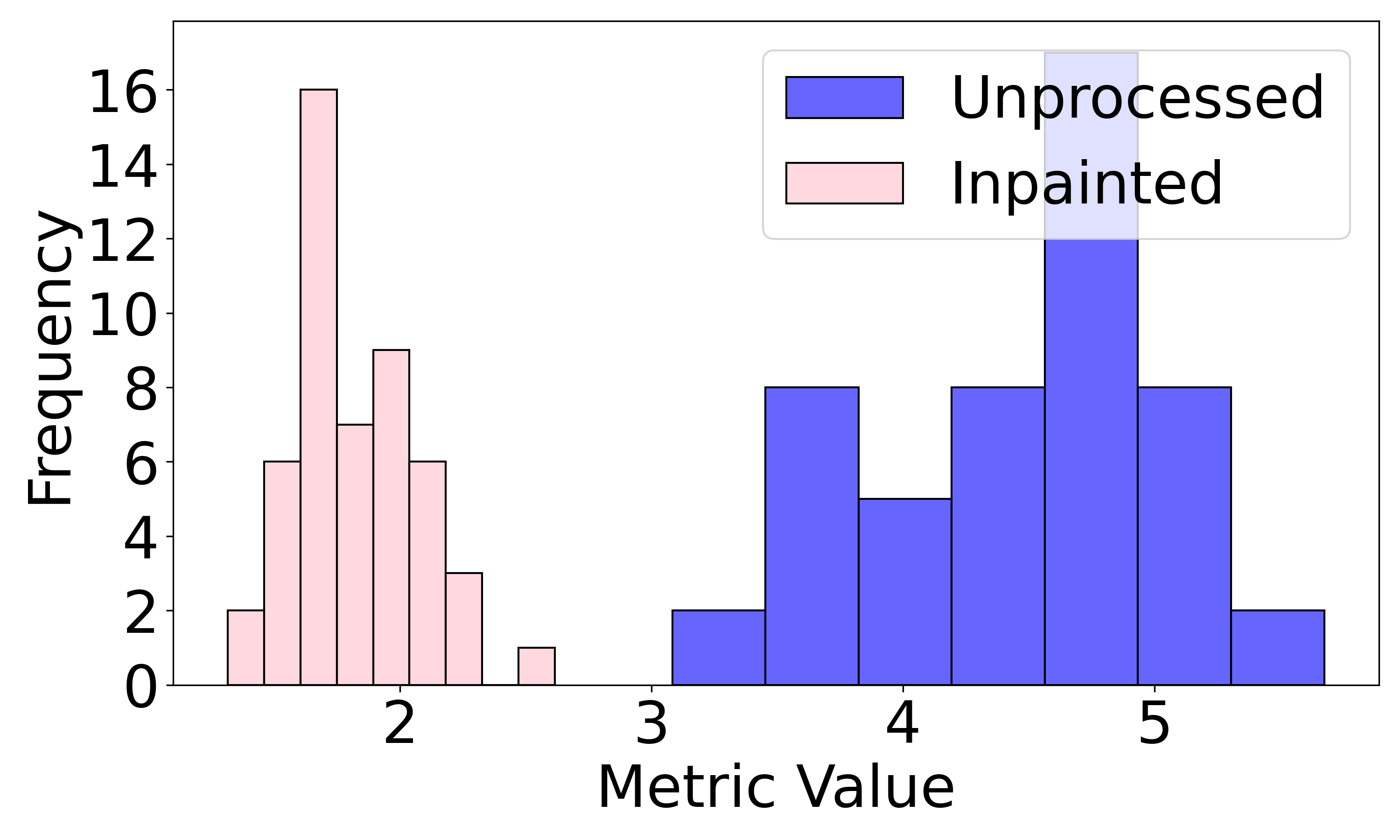} \\
            \tiny MCD = 4.74 $\xrightarrow{}$ 1.79
        \end{tabular} \\[0.5em]
        
        \rotatebox{90}{\hspace{0.8cm} \textsc{\footnotesize LogF0-MSE}} &
        \begin{tabular}[b]{@{}c@{}}
            \includegraphics[width=0.32\linewidth]{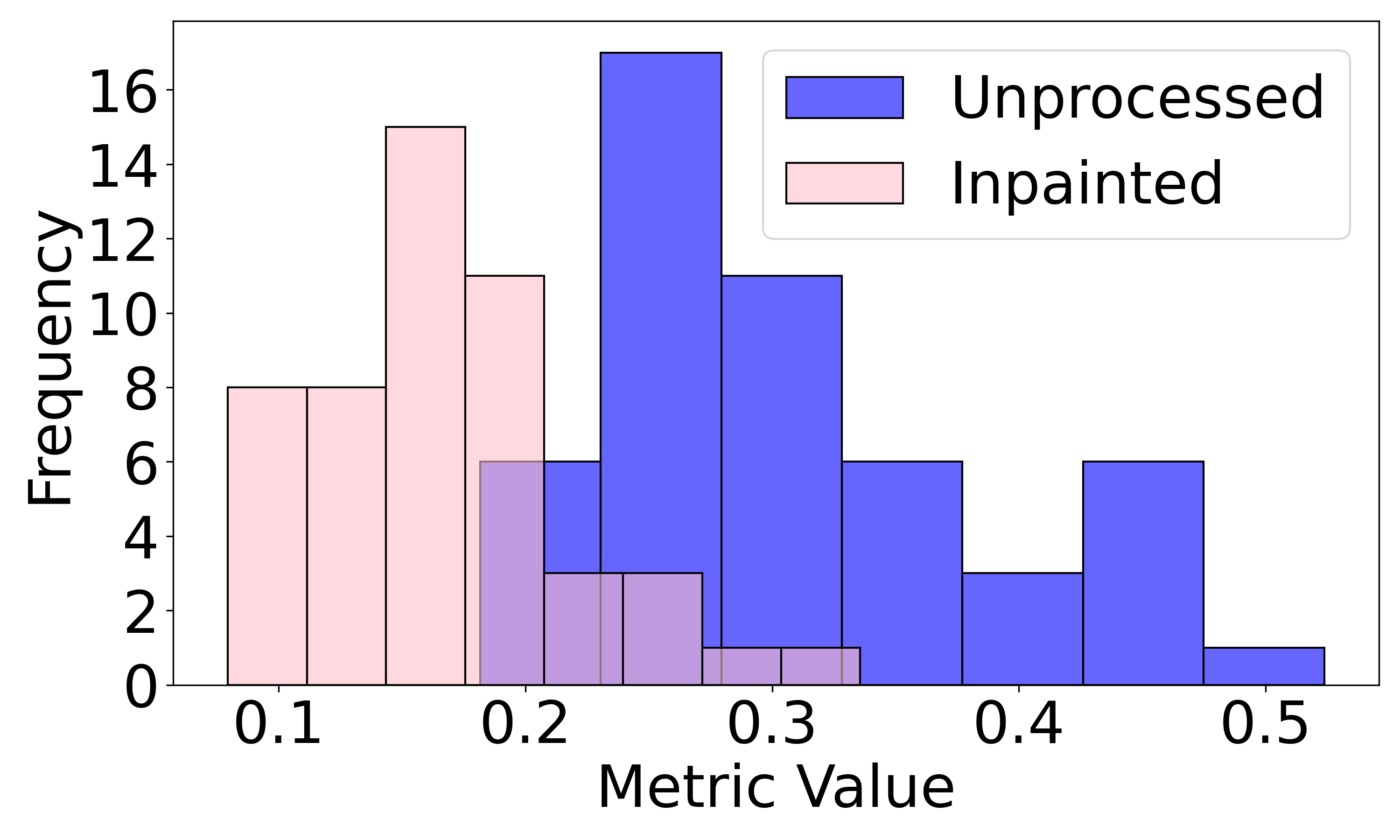} \\
            \tiny LogF0-MSE = 0.29 $\xrightarrow{}$ 0.17
        \end{tabular} &
        \begin{tabular}[b]{@{}c@{}}
            \includegraphics[width=0.32\linewidth]{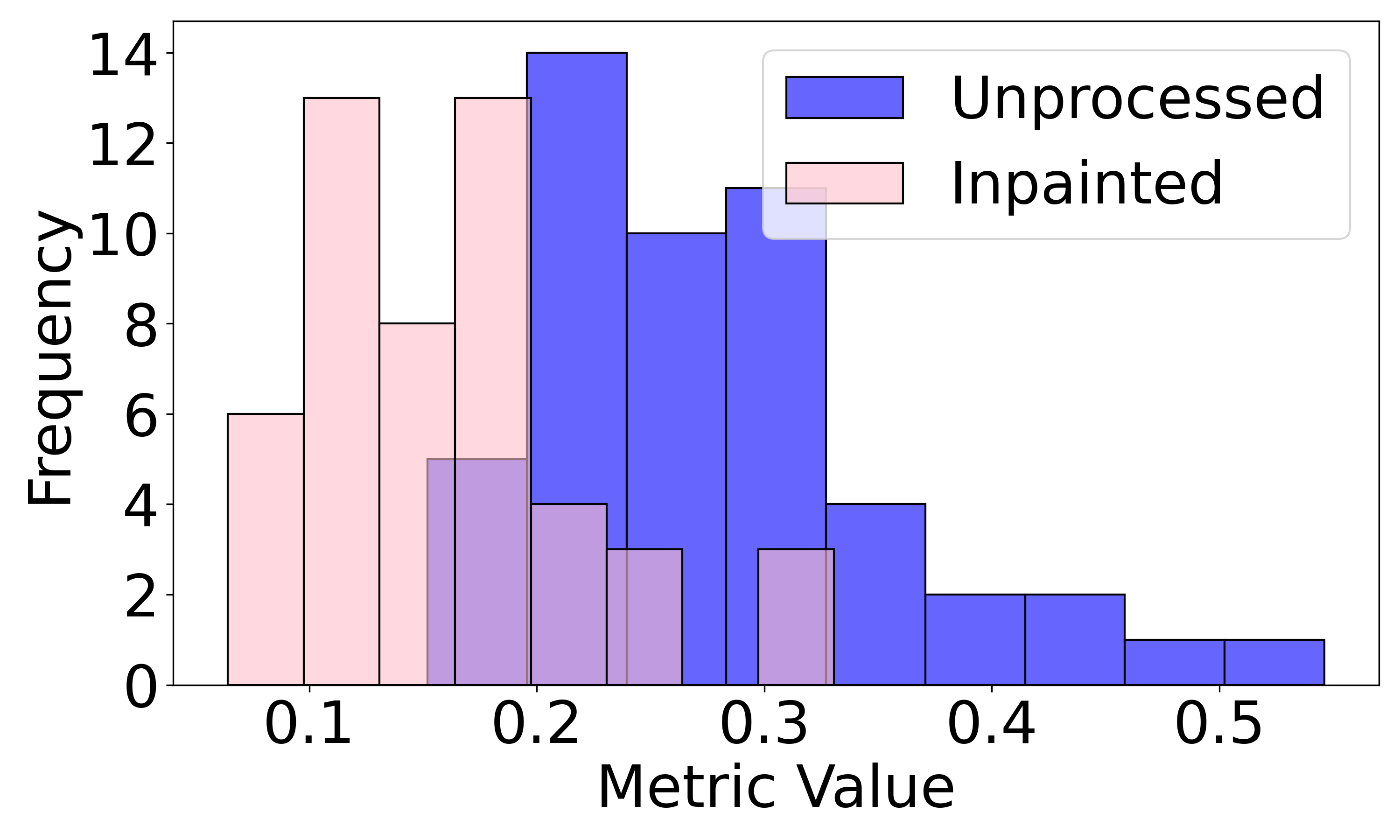} \\
            \tiny LogF0-MSE = 0.27 $\xrightarrow{}$ 0.15
        \end{tabular} &
        \begin{tabular}[b]{@{}c@{}}
            \includegraphics[width=0.32\linewidth]{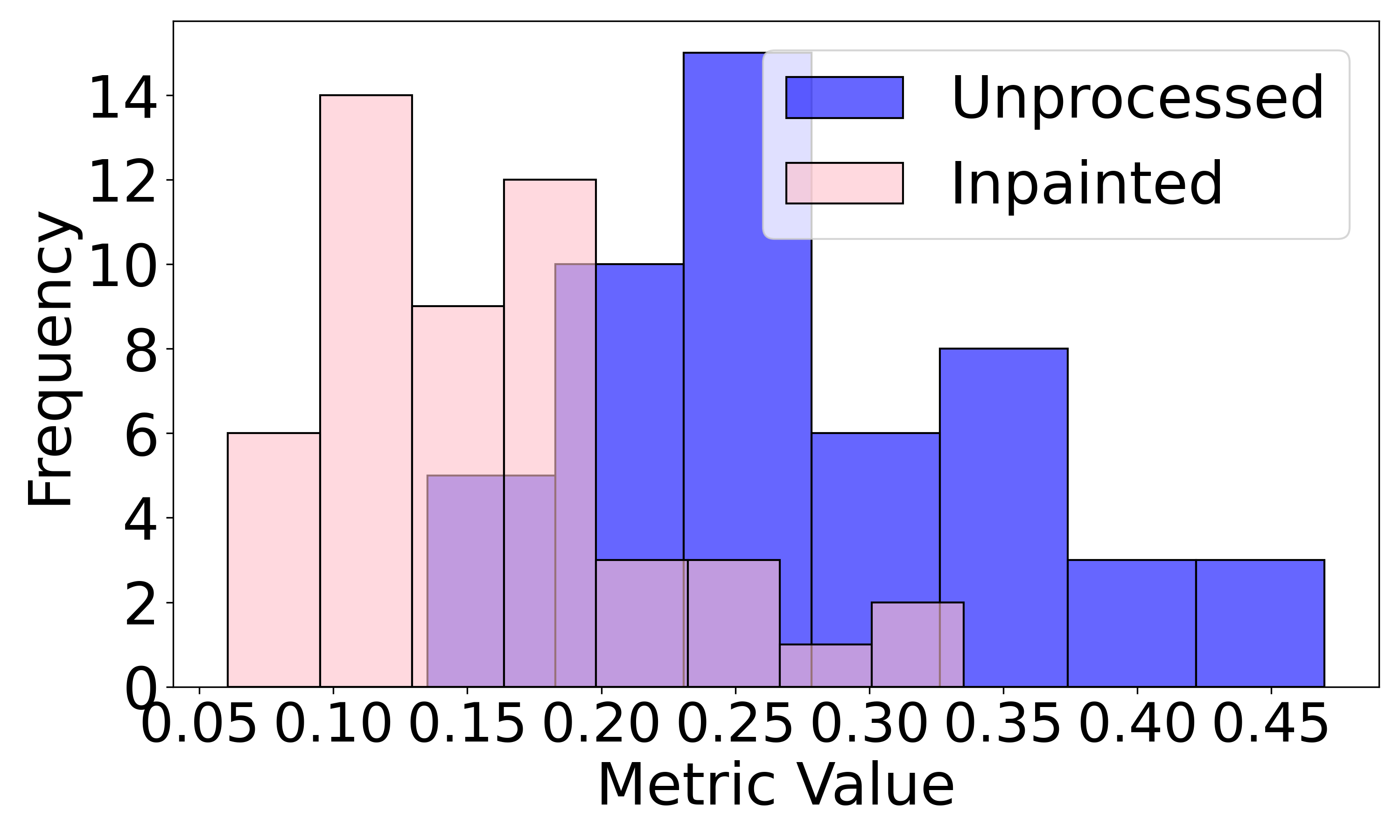} \\
            \tiny LogF0-MSE = 0.26 $\xrightarrow{}$ 0.15
        \end{tabular} \\
    \end{tabular}

    \caption{Comparison of evaluation metrics across gap durations. Columns show results for 0.1s, 0.2s, and 0.3s gaps. Rows show different metrics: WER, SpeechBLEU, Distance MOS, MCD, and LogF0-MSE. Values indicate median scores before and after inpainting.
    }\label{fig:gap_histograms_grid}
\end{figure}

In this section, we evaluate the performance of the proposed inpainting model when the transcript is not available but rather inferred from the masked signal using the ASR+LM module. The focus is on understanding how the model reconstructs speech using only the masked audio as input and how its performance varies with gap length. The evaluation includes multiple objective metrics and highlights both degradation patterns and strengths in prosody and acoustic consistency.
The results are presented as histograms in Fig.~\ref{fig:gap_histograms_grid}, illustrating the distribution of the evaluation metrics across different gap durations, as well as their median values.
First, across all evaluation metrics, the inpainted signals show consistent improvements compared to their initial unprocessed counterparts. Next, increasing the gap duration leads to higher \ac{WER} for both the unprocessed and inpainted signals because longer gaps make it more challenging to infer the correct transcription from the masked signal, as discussed in Section~\ref{exp:setup}. As a result, when the ground‑truth text is not provided, the overall performance degrades. Moreover, a comparison between Table~\ref{table:results_metric} and Fig.~\ref{fig:gap_histograms_grid} further illustrates this effect since the degradation stems from the ASR+LM system’s difficulty in predicting the correct transcript rather than from the inpainting model itself. Furthermore, the LogF0‑MSE values remain low, indicating that the inpainting process effectively preserves speaker pitch and, from the same arguments, the low MCD values demonstrate that the acoustic environment of the masked signal is maintained.

An interesting observation emerged when analyzing the results for the Distance MOS and SpeechBLEU metrics on the masked signal. Performance appears to improve as the gap duration increases, which initially seems counterintuitive. This can be explained by the fact that although the total duration of masking is held constant across different gap settings, the structure of the masking differs. Specifically, shorter gaps result in a greater number of brief masked segments, while longer gaps lead to fewer but more extended masked regions. Consequently, shorter gaps introduce more frequent interruptions in the signal, which may be more perceptually disruptive and lead to lower Distance MOS scores. For SpeechBLEU, which relies on 5-gram statistics, these frequent discontinuities associated with shorter gaps increase the mismatch between generated and reference tokens and result in lower scores compared to those obtained with longer, less frequent gaps.

\section{Conclusion}
\label{sec:conclusion}
In this work, we introduced \ac{PGDI}, a diffusion-based framework for speech inpainting under the assumption that the locations of the masked regions are known. When ground-truth text is available at inference time, we examined the impact of classifier guidance and found that phoneme-level conditioning significantly enhances reconstruction quality. Furthermore, we observed that the \ac{DiT} model consistently outperforms the U-Net architecture across all evaluation metrics and gap durations, particularly when guided by phonemes rather than word-level information. Based on these findings, we adopt the \ac{DiT} model with phoneme-based guidance for the inpainting task, even when the transcript is not available and must instead be inferred from the masked utterance using an ASR+LM module.

Notably, the diffusion-based model exhibits an inherent ability to preserve the prosody of the reconstructed speech even without any guidance, underscoring its capacity to model natural acoustic patterns. However, in the absence of guidance, semantic consistency is not maintained.

Our comprehensive experimental study shows that \ac{PGDI} can successfully reconstruct long missing segments, up to one second, while preserving the naturalness, prosody, and speaker identity. Although performance degrades for longer gaps compared to scenarios where the transcript is available, the pipeline remains effective even without access to the ground-truth transcript by predicting plausible phoneme sequences from the surrounding speech context.




\bibliographystyle{elsarticle-num}
\bibliography{Audio_Inpainting}

\end{document}

%% file: acronyms.tex
\acrodef{STFT}{Short-Time Fourier Transform}
\acrodef{ISTFT}{inverse short-time Fourier transform}
\acrodef{TF}{Time-Frequency}
\acrodef{RTF}{Relative Transfer Function }

\acrodef{LCMV}{Linearly Constrained Minimum Variance}

\acrodef{CSD}{Concurrent Speaker Detection}
\acrodef{VAD}{Voice Activity Detection}
\acrodef{OSD}{Overlapped Speech Detection}

\acrodef{NLP}{Natural Language Processing}

\acrodef{LSTM}{Long Short-Term Memory}
\acrodef{BLSTM}{Bidirectional Long Short-Term Memory}
\acrodef{RNN}{Recurrent Neural Networks}
\acrodef{CNN}{Convolutional Neural Networks}

\acrodef{GMM}{Gaussian mixture model}

\acrodef{ViT}{Vision Transformer}
\acrodef{CLS}{Class Token}
\acrodef{MHA}{multi-head attention}
\acrodef{TCN}{Temporal Convolutional Networks}

\acrodef{CE}{Cross-Entropy}
\acrodef{CS}{Cost-Sensitive}
\acrodef{LS}{Label-Smoothing}

\acrodef{mAP}{mean average precision}

\acrodef{SOTA}{state-of-the-art}
\acrodef{AST}{Audio Spectrogram Transformer}
\acrodef{ViT}{Vision Transformer}

\acrodef{CTC}{Connectionist Temporal Classification}
\acrodef{ASR}{Automatic Speech Recognition}
\acrodef{DiT}{Diffusion Transformer}
\acrodef{MSE}{mean squared error}
\acrodef{WER}{Word Error Rate}
\acrodef{STFT}{short-time Fourier transform}
\acrodef{DTFT}{Discrete Time Fourier Transform}
\acrodef{SSL}{self-supervised learning}
\acrodef{GMM}{Gaussian mixture models}
\acrodef{MCD}{Mel cepstral distortion}
\acrodef{NMR}{non-matching reference}
\acrodef{MOS}{mean opinion score}
\acrodef{BLEU}{bilingual evaluation understudy}
\acrodef{LogF0-MSE}{LogF0 Mean Squared Error}
\acrodef{TTS}{text to speech}
\acrodef{PGDI}{Phoneme-Guided Diffusion Inpainting}

\acrodef{SSL}{self-supervised learning}

\acrodef{LSTM}{long short-term memory}

\acrodef{GMM}{Gaussian mixture model}

\acrodef{LM}{language model}
\acrodef{DTW}{dynamic time warping}
\acrodef{adaLN-Zero}{zero-initialized adaptive Layer Norm}
\acrodef{adaLN}{adaptive Layer Norm}{}